\documentclass[onecolumn,showpacs,amsmath,amssymb]{revtex4}

\usepackage{amsmath}
\usepackage{graphicx}
\usepackage{dcolumn}
\usepackage{bm,color}

\begin{document}

\preprint{APS/}

\title{Energy density functional in nuclear physics}

\author{Yoritaka Iwata$^{1}$}
\email{y.iwata@gsi.de}
\author{Joachim A. Maruhn$^{2}$}
 \affiliation{$^{1}$GSI Helmholtzzentrum f\"ur Schwerionenforschung, D-64291 Darmstadt, Germany}
 \affiliation{$^{2}$Institut f\"ur Theoretische Physik, Universit\"at Frankfurt, D-60325 Frankfurt, Germany}


\begin{abstract}
Fundamentals of energy density functional in nuclear physics are presented.
Much attention is paid to a mathematically rigorous treatment of deriving the energy density functional.
The specific features of the density functional used in studying many-nucleon systems, which is quite different from that used in many-electron systems, are also shown.
The intended audience are physicists, chemists and mathematicians.
In particular those who will start to study the density functional theory are intended.
\end{abstract}

\pacs{21.60.Jz, 21.30.-x, 13.75.Cs}
\maketitle

\tableofcontents
\newpage

\section{Introduction} \label{sec0}
Among several methods in nuclear theory, the density functional method provides one of the most widely applicable treatments of many-nucleon systems.
For example, the nucleus is a many-nucleon system, where more than 300 stable nuclei are already known.
All atoms contain nuclei, and the chemical properties are identified by the number of protons included in the nucleus (cf. chemical elements: H, He, Li, $\cdots$).
It follows that the nucleus is one of the most important ingredients of our universe.

The theory based on the nucleonic degrees of freedom is presented, where the nucleon is not an elementary particle but consists of quarks and gluons.
The validity of this theory is based on the fact that nucleons are quite stable quantum entities that can be regarded as effective physical units.
Since nucleons are fermions, there are several common features relating this many-fermion system with other physical systems such as many-electron systems and so on.
However, there are several specific features in nuclear energy density functional.
Indeed, nucleons, which have isospin and spin degrees of freedom, interact by two completely different forces: nuclear and Coulomb forces.
In addition it is worth noting that reactions between nuclei are not similar to reactions of the other physical systems to a large extent.
Depending on the isospin degrees of freedom, there are two kinds of nucleons; i.e., protons and neutrons.
While we understand the Coulomb force well, much is not known about the nuclear force.
Indeed, although the quantum chromodynamics (QCD) Lagrangian has already been established, its connection to the nuclear force is still developing.
It makes many-nucleon system research quite difficult and also fascinating.

From a scientific point of view, many-nucleon research is associated with clarifying the origin, existence, structure and reaction of chemical elements, where the origin of elements heavier than iron has not been understood well.
It is an attempt to understand the time evolution of our universe with respect to the constituent chemical elements; ``How and where were all the chemical elements created and why do they exist as they are ?''. 
Chemical elements which do not naturally exist on earth are called superheavy elements.
These elements are artificially synthesized in the laboratory.
Such a superheavy element research is related with clarifying the existence limit of chemical elements.

Finite-body quantum systems are often investigated in the research of many-nucleon systems.
Although the spherical shape is expected to be energetically favoured if the system is governed simply by surface tension, the nucleus has experimentally been shown to have several shapes: e.g., spherical, prolate, oblate shapes and so on.
Such a research is associated with the structure of nuclei.

It is reasonable to have a unified theoretical framework describing both stationary and non-stationary states.
The nuclear density functional theory is a possible candidate. 
The basic equation of the many-nucleon system is the Schr{\"o}dinger equation containing the Hamiltonian.
Based on the independent particle motion, a possible form of the Hamiltonian is generally provided by 
\[  H = - \sum_{i=1}^A \frac{\hbar^2}{2m} \triangle_i + \sum \sum_{i<j} v_{i,j} +  \sum \sum \sum_{i<j<k} v_{ijk} \cdots. \]
The first term of the right hand side arises from the kinetic energy, and the other terms in the right hand side from the interaction energy.
For the interaction part, the density functional theory in many-nucleon system is a theory describing all the interaction by several densities (for many-electron systems, see Refs.~\cite{hohenberg,kohn}).
As is already mentioned, the interaction part of the Hamiltonian is not perfectly known as far as the many-nucleon system is concerned, so that the one important task in nuclear density functional theory is to find out the ultimate interaction, which can describe all the phenomena in the many-nucleon system.
This point is indeed different from studying many-electron systems.
The process of finding the ultimate effective Hamiltonian mainly consists of two steps;
first, to find an appropriate functional form of the effective Hamiltonian; second, to find its best parameter sets. 

In this chapter much attention is paid to show a method of deriving the effective interaction in many-nucleon systems.
What is presented in this chapter is a kind of modelling: the modelling of interacting femtometer-scale fermions.
The nucleon-nucleon interaction is decomposed in Sec.~\ref{sec1}.
The effective Hamiltonian arising from the nuclear force is discussed in Sec.~\ref{sec2}; starting from the zero-range nucleon-nucleon interaction (Sec.~\ref{sub1}), we show a procedure of obtaining the Hamiltonian density (Sec.~\ref{sub2}); the effective Hamiltonian is obtained by applying the variational principle~ (Sec.\ref{sub3}).
The effective Hamiltonian arising from additional forces is briefly discussed in Sec.~\ref{sec3}.
In particular we provide a mathematically rigorous treatment of applying the variational principle, where infinite-dimensional Hilbert/Banach spaces are considered.
For the mathematics used in this chapter, refer to the textbooks of functional analysis such as Ref.~\cite{yosida}.   \vspace{16mm} \\

\section{Nucleon-nucleon interaction} \label{sec1}
There are two kinds of nucleons (isospin degree of freedom $\tau$): protons and neutrons, which are fermions.
In addition the nucleon has the spin degree of freedom.
Therefore we are interested in the many-fermion system with spin and isospin degrees of freedom.

Nucleons interact by two kinds of forces: nucleon-nucleon interaction $V$ is represented by the sum of the nuclear potential $V_K$, the Coulomb potential $V_C$ and the pairing potential $V_{\rm pair}$: 
\begin{equation} V = V_K + V_C + V_{\rm pair}. \end{equation}
Only protons interact by the Coulomb force.
The two kinds of nucleons form a bound system called atomic nucleus.
The atomic nucleus is a finite quantum system containing $Z$ protons and $N$ neutrons, where $A=Z+N$ is called mass number.
One of the goals of theory is to determine for which combinations $Z$, $N$ can exist.  \vspace{16mm} \\

\section{Nuclear force} \label{sec2}
Let us begin with the primitive picture of the nuclear force using the potential description.
The potential of nuclear force depends on the positions, momenta, spins (${\bm \sigma}_i = \pm \frac{1}{2}$) and isospins (${\bm \tau}_i = \pm \frac{1}{2}$) of the two nucleons:
\[ V_K = V_K({\bm r}_i,{\bm r}_j,{\bm p}_i,{\bm p}_j,{\bm \sigma}_i,{\bm \sigma}_j,{\bm \tau}_i,{\bm \tau}_j),  \]
where $i$ and $j$ are the indices identifying the two nucleons.
It is rational to determine interactions obeying invariance, because the physical law must be invariant regardless of observers.
The following invariance are required for $V_K$: translational invariance, Galilean invariance, rotational invariance, isospin invariance, parity invariance, and time-reversal invariance (for each invariance, see Sec. 7.1.1 of Ref.~\cite{Greiner-Maruhn}). 
It is worth noting here that the radial dependence of the function $V_K$ cannot be deduced from invariance principles.
Among several attempts of determining the radial dependence, H. Yukawa proposed the Yukawa potential:
\[ V_{Y}(r)  = \frac{e^{-\mu \gamma}}{\mu \gamma}  \]
based on meson field theory, where $1/\mu$ is the Compton wavelength of the pion.

Historically the components of the nuclear force are represented using the identity operator, $({\bm \sigma}_i \cdot {\bm \sigma}_j)$, $( {\bm \tau}_i \cdot {\bm \tau}_j)$ and $({\bm \sigma}_i \cdot {\bm \sigma}_j)({\bm \tau}_i \cdot {\bm \tau}_j)$ multiplied by the spin- and isospin- independent ingredient.
In the traditional formulation, such a nucleon-nucleon interaction is represented using several exchange operators:
\begin{equation} \label{trad-int}  \begin{array}{ll} 
V_K = V_{\rm W}({\bm r}) + V_{\rm M}({\bm r}) P_r + V_{\rm B}({\bm r}) P_{\sigma}  - V_{\rm H}({\bm r}) P_{\tau} \vspace{1.5mm}  \\
\quad  = V_{\rm W}({\bm r}) + V_{\rm M}({\bm r}) P_r + V_{\rm B}({\bm r}) P_{\sigma}  + V_{\rm H}({\bm r}) P_{r} P_{\sigma},  
\end{array} \end{equation}
where $P_r$, $P_{\sigma} = \frac{1}{2} (1+ {\bm \sigma}_i {\bm \sigma}_j)$ and $P_{\tau} =  \frac{1}{2} (1+ {\bm \tau}_i {\bm \tau}_j)$ are exchange operators of coordinates, spins and isospins.
In particular the fermionic relation $ P_{r} P_{\sigma} P_{\tau} = -1$ is used to replace the last term, and every term is represented without $P_{\tau}$ (for the property of fermions, refer to textbooks of quantum physics such as Ref.~\cite{fetter}).
The indices of the first, second, third and fourth terms stand for Wigner, Majorana, Bartlett and Heisenberg, respectively.
Another important ingredient is the tensor force represented by $({\bm r}_i \cdot {\bm \sigma}_i) ({\bm r}_i \cdot {\bm \sigma}_j)$, where the tensor operator is
\[
S_{ij} = \frac{3 ({\bm r}_i \cdot {\bm \sigma}_i) ({\bm r}_i \cdot {\bm \sigma}_j)}{|{\bm r}_i|^2} -({\bm \sigma}_i \cdot {\bm \sigma}_j).
 \]
The aim of this chapter is to have a density functional representation for the nuclear force (for a whole process, see Fig.~1), which includes the exchange properties represented by $({\bm \sigma}_i \cdot {\bm \sigma}_j)$, $( {\bm \tau}_i \cdot {\bm \tau}_j)$, $({\bm \sigma}_i \cdot {\bm \sigma}_j)({\bm \tau}_i \cdot {\bm \tau}_j)$ and $({\bm r}_i \cdot {\bm \sigma}_i) ({\bm r}_i \cdot {\bm \sigma}_j)$.
The interaction described by spin and isospin degrees of freedom is specific to nuclear physics.  
\vspace{16mm} \\

\begin{figure} \label{fig1} 
\includegraphics[width=12cm]{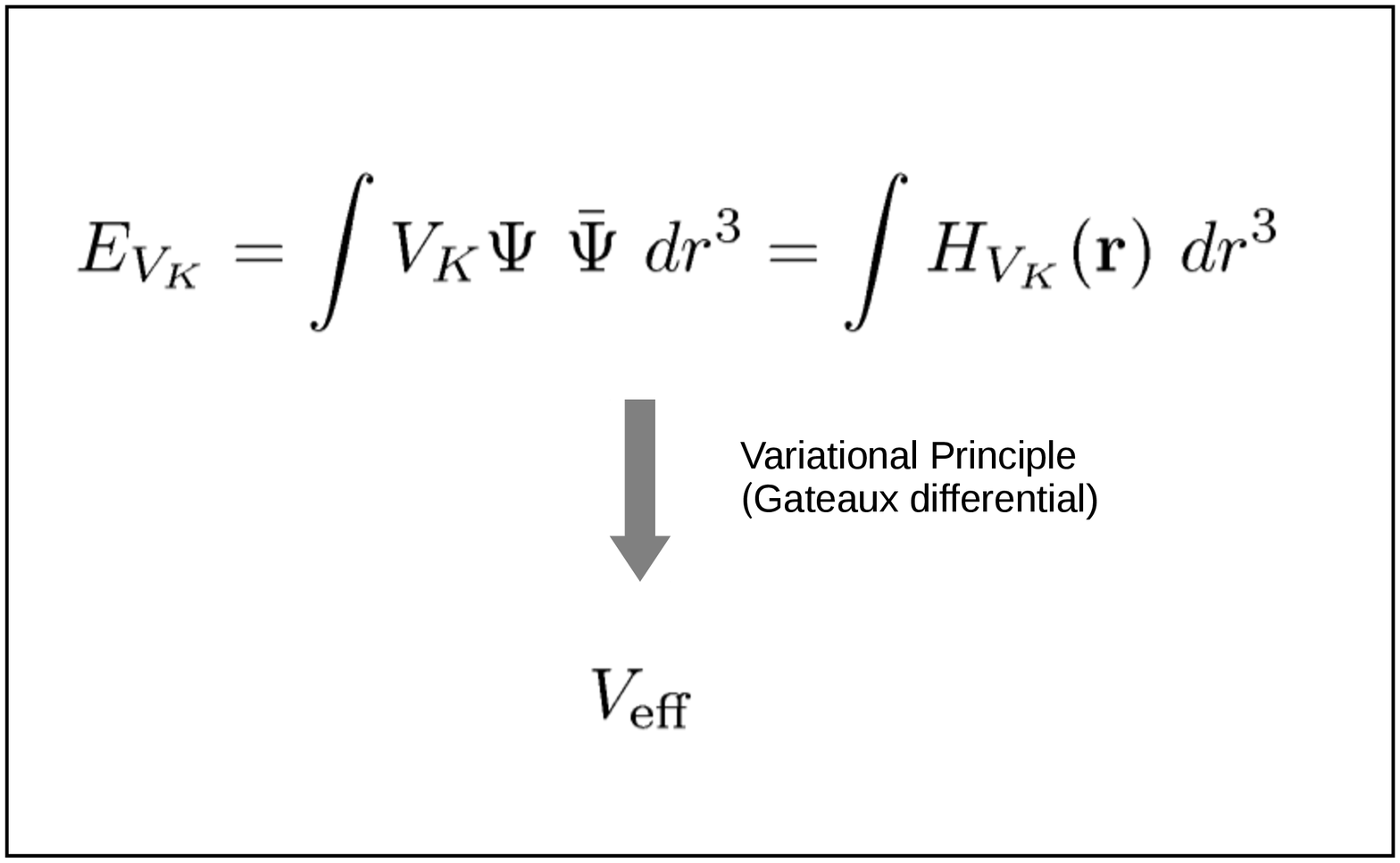} \\
\caption{A process of obtaining the effective interaction.
$E_{V_K}$ denotes the interaction energy, $V_K$ is the zero-range interaction, $\Psi$ denotes a trial function, and $H_{V_K}({\rm r})$ is the interaction part of the Hamiltonian density.
$V_{\rm eff}$, which is obtained by calculating the Gateaux differential, is the effective interaction in the density functional formalism.
The zero-range interaction $V_K$ is presented in Sec.~\ref{sub1}, the energy and the Hamiltonian density are calculated in Sec.~\ref{sub2}, and the effective interaction is obtained in Sec.\ref{sub3}.
}
\end{figure}

\subsection{The zero-range force formalism \label{sub1}}

For the purpose of obtaining the energy density functional in many-nucleon system, it is reasonable to begin with the nucleon-nucleon interaction of the following form: 
\begin{equation} \label{v-rep} \displaystyle  V_{K} = {\sum \sum }_{i<j} v_{i,j} +  {\sum \sum \sum}_{i<j<k} v_{i,j,k} \cdots,  \end{equation}
where the first and second terms of the right hand side are the two-body (two-nucleon) and three-body (three-nucleon) interactions, respectively.
In this formalism all the interactions between each nucleon are summed up. 
With respect to the interaction used in density functional calculations, nucleon-nucleon interaction is usually considered up to the three-body term (cf. Skyrme interaction~\cite{skyrme}). 

In this context special treatment (the short range approximation) is necessary to represent non-local potentials. 
First, the relative momentum between two nucleons, which is necessary to describe the interaction, is assumed to be represented by
\begin{equation} \label{delta} {\bm k} = \frac{1}{2i} (\nabla_i - \nabla_j), \quad {\bm k}' = - \frac{1}{2i} (\nabla_i' - \nabla_j'),   \end{equation}
where, again, $i$ and $j$ are the indices identifying the two nucleons. 
${\bm k}'$ acts on the wave function to its left.
For this representation, it is useful to remember the Einstein-de Broglie formula: ${\bm p} = \hbar {\bm k}$ under the quantum mechanical correspondence: ${\bm p} \to (\hbar/i)  \nabla$.
This treatment is essential to obtain the zero-range interaction.

The two-body interaction, which contains momentum dependence as well as spin-exchange contributions, is derived by assuming the polynomial expansion in powers of ${\bm k}$ and ${\bm k}'$ with the low-momentum cut-off up to quadratic terms (linear terms in ${\bm k}$ are forbidden by time-reversal invariance):
\begin{equation} \label{skyrm}  \begin{array}{ll}
 v_{i,j}({\bm k},{\bm k}')  = t_0 (1 + x_0 P_{\sigma})  \delta({\bm r}_i - {\bm r}_j)) 
       + \frac{t_1}{2} (1 + x_1 P_{\sigma}) \{ \delta({\bm r}_i - {\bm r}_j) {\bm k}^2 + {\bm k}'^2  \delta({\bm r}_i - {\bm r}_j) \}  
 + t_2 (1 + x_2 P_{\sigma}) {\bm k}'  \delta({\bm r}_i - {\bm r}_j)  {\bm k}   \vspace{1.5mm}  \\
\quad  + \frac{t_e}{2} [ \{ 3 ({\bm \sigma}_i \cdot {\bm k}')( {\bm \sigma}_j \cdot {\bm k}') - ({\bm \sigma}_i \cdot {\bm \sigma}_j) {\bm k}'^2 \}  \delta({\bm r}_i - {\bm r}_j)
 + \delta({\bm r}_i - {\bm r}_j) \{ 3 ({\bm \sigma}_i \cdot {\bm k})( {\bm \sigma}_j \cdot {\bm k}) - ({\bm \sigma}_i \cdot {\bm \sigma}_j) {\bm k}^2 \} ]   \vspace{1.5mm}  \\
\quad  + \frac{ t_o }{2} [3 ({\bm \sigma}_i \cdot {\bm k}') \delta({\bm r}_i - {\bm r}_j) ( {\bm \sigma}_j \cdot {\bm k}) - ({\bm \sigma}_i \cdot {\bm \sigma}_j) {\bm k}'  \delta({\bm r}_i - {\bm r}_j) {\bm k}],  
\end{array} \end{equation}
where $t_0$, $t_1$, $t_2$, $t_e$ and $t_o$ are parameters (following the traditional manner of representation), $P_{\sigma} = \frac{1}{2} (1+ {\bm \sigma}_i \cdot {\bm \sigma}_j) $ is the spin-exchange operator, and ${\bm \sigma}_i$ are the Pauli matrices.
The quadratic cut-off implies that the interaction shown in Eq~\eqref{skyrm} is valid to low-energy situations (cf. the fermi energy).
The $t_0$ term represents the central potential, the $t_1$ and $t_2$ terms stand for the non-local potential, and the $t_o$ and $t_e$ terms are derived from the tensor interaction.
It is important to introduce the spin-orbit interaction, which is an additional effect necessary to reproduce the nuclear structure.
\begin{equation} \begin{array}{ll}
 v^{LS}_{i,j}({\bm k},{\bm k}') =   i W_0 ~ {\bm k}' ~ \delta({\bm r}_i - {\bm r}_j)~ (\sigma_i + \sigma_j) \times {\bm k}.   
\end{array} \end{equation}
Another important ingredient is a two-body interaction with density dependence:
\begin{equation} \begin{array}{ll}
v^{DD}_{i,j}({\bm k},{\bm k}')  = \frac{t_3}{6} (1+ x_3 P_\sigma )  \rho(\frac{{\bm r}_i + {\bm r}_j}{2})^{\alpha}  \delta({\bm r}_i - {\bm r}_j). 
\end{array} \end{equation}
This provides a treatment of medium effects including three-body and many-body forces.
Note that the introduction of this force is due to the phenomenological reason.

As a result, the nuclear interaction is represented by the zero-range Skyrme interaction:
\begin{equation} \label{Skyrme}
V_{i,j}({\bm k},{\bm k}') = v_{i,j}({\bm k},{\bm k}') +  v^{LS}_{i,j}({\bm k},{\bm k}') +  {v^{DD}_{i,j}}({\bm k},{\bm k}'),
\end{equation}
where note that the included momenta ${\bm k}$ and ${\bm k}'$ are represented by Eq. (\ref{delta}). 
 \vspace{16mm} \\

\subsection{Hamiltonian density  \label{sub2}}
The Hamiltonian density is obtained by calculating the energy expectation value from the zero-range nuclear interaction \eqref{Skyrme}.
The following points, which are illustrated in concrete discussion, should be noticed:
\begin{itemize}
\item Fermionic relation is assumed in deriving the Hamiltonian density.
\item The boundary condition is necessary.
\item Differential operators appear in deriving non-local interaction part of the Hamiltonian density.
\end{itemize}
In this section (Sec.~\ref{sub2}), wave functions are assumed to be sufficiently smooth. 
With respect to the many-nucleon system as a fermionic many-body system, fermionic relation ($P_r P_{\sigma} P_{\tau} = -1$), which has already mentioned in Sec.~\ref{sub1}, has to be taken into account:
\begin{equation} \label{asym}  V_{i,j}({\bm k},{\bm k}') = \frac{1}{2}  V_{i,j}({\bm k},{\bm k}') (1- P_r P_{\sigma} P_{\tau}). \end{equation}
Using Eqs. \eqref{Skyrme} and \eqref{asym}, the energy is obtained by calculating
\begin{equation} \label{energy}  \begin{array}{ll}
E = E_{t_0} +  E_{t_1} +  E_{t_2} +  E_{t_3} +  E_{t_e} +  E_{t_o} +  E_{W_0}  \vspace{1.5mm}\\
  = \frac{1}{2} \sum_{l,m} \int {\bar \psi}_l ({\bm r}_i') {\bar \psi}_m({\bm r}_j')  V_{i,j}({\bm k},{\bm k}') (1- P_r P_{\sigma} P_{\tau})  \psi_l ({\bm r}_i) \psi_m({\bm r}_j) d{\bm r}_i d{\bm r}_j d{\bm r}_i' d{\bm r}_j',  
\end{array} \end{equation}
where $E_{t_0}$, $E_{t_1}$ $\cdots$, $E_{W_0}$ denote the energy arising from the terms with the coefficient $t_0$, $t_1$ $\cdots$, $W_0$, respectively.
In Eq.~\eqref{energy}, $\psi_l ({\bm r}_i) \psi_m({\bm r}_j)$ plays a role of trial function including two single-particle wave functions.
Furthermore we assume that there is no isospin mixing in the nuclear force.
Since the expectation value $E$ of the many-body Hamiltonian can also be written using Hamiltonian density $H({\bm r})$:
\begin{equation}
 E  = \int dr^3 H({\bm r}), 
\end{equation}
we obtain the representation for the Hamiltonian density.
Among several terms, in order to see some principal mathematical features (e.g., the appearance of differential operators) of the density functional formalism in nuclear physics, we examine the following energies, which are picked out from Eq.~\eqref{energy}.
\[ E = E_{t_0} +  E_{t_1} +  E_{t_2} +  E_{t_3},  \]
where this choice is also due to the pedagogical reasons.

Let $\psi$ be a single wave function, and introduce the densities.
Before calculating the Hamiltonian density it is necessary to introduce the following densities.
The nonlocal particle density and the nonlocal spin density are defined by
\[ \begin{array}{ll}
 \rho ({\bm r},{\bm r}') =  \sum_{\sigma,q} \rho_{q} ({\bm r},{\bm r}',\sigma)  =  \sum_{l,\sigma,q} {\bar \psi}_l ({\bm r}',\sigma,q) \psi_l({\bm r},\sigma,q), \vspace{2.5mm} \\
{\bm s}({\bm r},{\bm r}')  =  \sum_{\sigma,\sigma', q} \rho_{q} ({\bm r},{\bm r}',\sigma,\sigma') <\sigma'|{\bm \sigma}|\sigma>  = \sum_{l,\sigma,\sigma',q} {\bar \psi}_l({\bm r}',\sigma',q) ~{\bm \sigma}~  \psi_l({\bm r},\sigma,q) 
\end{array} \]
using positions ${\bm r}$ and ${\bm r}'$, isospin $q$, and spins $\sigma$ and $\sigma'$.
According to Engel et al.~\cite{engel-brink}, local densities are defined by
\begin{equation} \label{density} \begin{array}{ll}
\rho ({\bm r}) = \rho ({\bm r},{\bm r}')|_{{\bm r} = {\bm r}'},  \vspace{2.5mm}  \\
\tau ({\bm r}) = \nabla \cdot \nabla'  \rho ({\bm r},{\bm r}')|_{{\bm r} = {\bm r}'},  \vspace{2.5mm} \\
s_{\mu}({\bm r}) =  s_{\mu} ({\bm r},{\bm r}')|_{{\bm r} = {\bm r}'},   \vspace{2.5mm} \\
j_{\mu} ({\bm r}) = -\frac{i}{2}(\nabla_{\mu} - \nabla'_{\mu}) \rho ({\bm r},{\bm r}')|_{{\bm r} = {\bm r}'},  \vspace{2.5mm}  \\
J_{\mu \nu} ({\bm r}) = -\frac{i}{2}(\nabla_{\mu} - \nabla'_{\mu}) s_{\nu} ({\bm r},{\bm r}')|_{{\bm r} = {\bm r}'},  \vspace{2.5mm}  \\
T_{\mu} ({\bm r}) =  \nabla \cdot \nabla'  s_{\mu} ({\bm r},{\bm r}')|_{{\bm r} = {\bm r}'}, 
\end{array} \end{equation}
where the indices $\mu$ and $\nu$ indicate the Cartesian coordinate components $x$, $y$, and $z$.
In particular $\mu \nu$ in $J_{\mu \nu}({\bm r})$ can take only unequal values to $\mu$ and $\nu$ (cf. exterior product).
The first two densities, that is, the particle density $\rho({\bm r})$ and the kinetic energy density $\tau({\bm r})$ are scalar quantities, while the others, the spin density $s_{\mu}({\bm r})$, the current density $j_{\mu} ({\bm r})$, the spin-current density $J_{\mu \nu} ({\bm r})$, and the spin kinetic energy density $T_{\mu} ({\bm r})$ are vector quantities.  
The index $q$ for isospin is used instead of $\tau$ because it is used for the kinetic energy density.

Galilean invariance restricts the form of the Hamiltonian density~\cite{engel-brink}.
Galilean transformation replaces a single-particle wave function $\psi({\bm r}, \sigma,q)$ by $e^{i{\bm k} {\bm r}} \psi ({\bm r}, \sigma,q)$.
Thus the densities transform as
\[ \begin{array}{ll}
 \rho({\bm r},{\bm r}') ~\to~ e^{i{\bm k} \cdot ({\bm r}- {\bm r}')} \rho({\bm r},{\bm r}'),  \vspace{1.5mm}  \\
 \tau({\bm r}) ~\to~ \tau({\bm r}) + 2{\bm k}\cdot {\bm j}({\bm r})+{\bm k}^2 \rho({\bm r}), \vspace{1.5mm} \\
 {\bm j}({\bm r}) ~\to~ {\bm j}({\bm r})+ {\bm k} \rho({\bm r}), 
\end{array} \]
because 
\[ \begin{array}{ll}  e^{-i{\bm k}{\bm r}'} {\bar \psi}({\bm r'}, \sigma,q)~ e^{i{\bm k}{\bm r}} \psi ({\bm r}, \sigma,q) = e^{i{\bm k}({\bm r}-{\bm r}')} {\bar \psi} ({\bm r'}, \sigma,q)~ \psi ({\bm r}, \sigma,q),  \vspace{5mm}  \\
 (\nabla e^{-i{\bm k}{\bm r}'} {\bar \psi} ({\bm r'}, \sigma,q))~ (\nabla e^{i{\bm k}{\bm r}} \psi ({\bm r}, \sigma,q))   \vspace{1.5mm}  \\
\quad = ( -i{\bm k}e^{-i{\bm k}{\bm r}'} {\bar \psi} ({\bm r'}, \sigma,q) + e^{-i{\bm k}{\bm r}'} \nabla {\bar \psi} ({\bm r'}, \sigma,q))~ 
(i{\bm k}e^{i{\bm k}{\bm r}} \psi ({\bm r}, \sigma,q) + e^{i{\bm k}{\bm r}} \nabla \psi ({\bm r}, \sigma,q))  \vspace{1.5mm}  \\
\quad = {\bm k}^2 e^{i{\bm k}({\bm r}-{\bm r}')} {\bar \psi} ({\bm r'}, \sigma,q) \psi ({\bm r}, \sigma,q) +  e^{i{\bm k}({\bm r}-{\bm r}')} \nabla {\bar \psi} ({\bm r'}, \sigma,q) \nabla \psi ({\bm r}, \sigma,q)  \vspace{1.5mm}  \\
\qquad  -i{\bm k}e^{i{\bm k}({\bm r}-{\bm r}')} {\bar \psi} ({\bm r'}, \sigma,q)~(\nabla \psi ({\bm r}, \sigma,q))
 +i{\bm k}e^{i{\bm k}({\bm r}-{\bm r}')} (\nabla {\bar \psi} ({\bm r'}, \sigma,q))~ \psi ({\bm r}, \sigma,q),  \vspace{5mm}  \\ 
 -(i/2) \{ e^{-i{\bm k}{\bm r}'} {\bar \psi} ({\bm r'}, \sigma,q)~(\nabla  e^{i{\bm k}{\bm r}}  \psi ({\bm r}, \sigma,q))
 -  (\nabla e^{-i{\bm k}{\bm r}'} {\bar \psi} ({\bm r'}, \sigma,q))~  e^{i{\bm k}{\bm r}} \psi ({\bm r}, \sigma,q) \}   \vspace{1.5mm}  \\
\quad = -(i/2)\{ e^{-i{\bm k}{\bm r}'} {\bar \psi}({\bm r'}, \sigma,q)~(i{\bm k} e^{i{\bm k}{\bm r}}  \psi ({\bm r}, \sigma,q) +  e^{i{\bm k}{\bm r}} \nabla \psi ({\bm r}, \sigma,q) )  \vspace{1.5mm}  \\
\qquad -  ( -i{\bm k} e^{-i{\bm k}{\bm r}'} {\bar \psi} ({\bm r'}, \sigma,q) + e^{-i{\bm k}{\bm r}'} \nabla {\bar \psi} ({\bm r'}, \sigma,q))~  e^{i{\bm k}{\bm r}} \psi ({\bm r}, \sigma,q) \}    \vspace{1.5mm}  \\
\quad = -(i/2) \{e^{i{\bm k}({\bm r}-{\bm r}')}  {\bar \psi} ({\bm r'}, \sigma,q) \nabla \psi ({\bm r}, \sigma,q)  -  e^{i{\bm k}({\bm r}-{\bm r}')} \nabla {\bar \psi} ({\bm r'}, \sigma,q) \psi ({\bm r}, \sigma,q) 
 +2 i{\bm k} e^{i{\bm k}({\bm r}-{\bm r}')} {\bar \psi} ({\bm r'}, \sigma,q)~ \psi ({\bm r}, \sigma,q)   \}.  
\end{array} \]
The transformation leaves $\rho \tau - {\bm j}^2$ invariant. Indeed, $e^{ik \cdot ({\bm r}- {\bm r}')} \rho({\bm r},{\bm r}')|_{{\bm r}={\bm r}'} = \rho({\bm r}) $ leads to
\[ \begin{array}{ll}
 \rho ( \tau + 2{\bm k}\cdot {\bm j}+{\bm k}^2 \rho ) - ({\bm j}+ {\bm k}\rho)^2 
=  \rho ( \tau + 2{\bm k}\cdot {\bm j}+{\bm k}^2  \rho) - {\bm j}^2 - 2 {\bm j} \cdot {\bm k} \cdot \rho -  {\bm k}^2 \rho^2 
=  \rho \tau  - {\bm j}^2. 
\end{array} \]
Consequently $ {\bm j}^2$ cannot appear alone and the corresponding ingredient of the Hamiltonian density is $\rho \tau - {\bm j}^2$ instead.

We use the following substitution through integration by parts:
\begin{equation}  \begin{array}{ll}
 (\nabla  \rho({\bm r}))^2 
 ~\to~ - \rho({\bm r}) ~ (\nabla^2  \rho({\bm r})),
\end{array} \end{equation}
which can be valid under a suitable boundary condition.
In this section (Sec.~\ref{sub2}), we assume that the Dirichlet-zero boundary condition is imposed.
Note that nonzero additional boundary integral remains if we have no boundary condition.
We have
\begin{equation} \label{tools}  \begin{array}{ll}
 \nabla  \rho({\bm r},{\bm r}')|_{{\bm r}={\bm r}'} 
 =  \sum_{l,\sigma,q} {\bar \psi}_l ({\bm r}',\sigma,q) \nabla \psi_l({\bm r},\sigma,q) |_{{\bm r}={\bm r}'}
 =  \sum_{l,\sigma,q} {\bar \psi}_l ({\bm r},\sigma,q) \nabla \psi_l({\bm r},\sigma,q)  \vspace{1.5mm}\\  
\quad  =  \sum_{l,\sigma,q} \{ \frac{1}{2}( {\bar \psi}_l ({\bm r},\sigma,q) \nabla \psi_l({\bm r},\sigma,q ) +  \psi_l ({\bm r},\sigma,q) \nabla {\bar \psi}_l({\bm r},\sigma,q) ) \vspace{1.5mm}\\
\qquad + \frac{1}{2}( {\bar \psi}_l ({\bm r},\sigma,q) \nabla \psi_l({\bm r},\sigma,q) -  \psi_l ({\bm r},\sigma,q) \nabla {\bar \psi}_l({\bm r},\sigma,q) )  \}  \vspace{1.5mm}\\     
\quad = \frac{1}{2} \nabla  \rho({\bm r}) +  i{\bm j}({\bm r}), \vspace{5mm}\\
 \nabla'  \rho({\bm r},{\bm r}')|_{{\bm r}={\bm r}'}  = \frac{1}{2} \nabla \rho({\bm r}) -  i{\bm j}({\bm r}),\vspace{5mm}\\
(\nabla^2 +  \nabla'^2) \rho({\bm r},{\bm r}')|_{{\bm r}={\bm r}'}  =  \sum_{l,\sigma,q} {\bar \psi}_l ({\bm r}',\sigma,q) \nabla^2 \psi_l({\bm r},\sigma,q) + \psi_l ({\bm r},\sigma,q) \nabla'^2 {\bar \psi}_l({\bm r}',\sigma,q) |_{{\bm r}={\bm r}'}  \vspace{1.5mm}\\ 
\quad =  \sum_{l,\sigma,q} \{ ( {\bar \psi}_l ({\bm r}',\sigma,q) \nabla^2 \psi_l({\bm r},\sigma,q) + \psi_l ({\bm r},\sigma,q) \nabla'^2 {\bar \psi}_l({\bm r}',\sigma,q)   + 2  \nabla' {\bar \psi}_l({\bm r}',\sigma,q)  \nabla \psi_l({\bm r},\sigma,q))   \vspace{1.5mm}\\
\qquad  - 2 \nabla' {\bar \psi}_l({\bm r}',\sigma,q)  \nabla \psi_l({\bm r},\sigma,q)   \} )|_{{\bm r}={\bm r}'}  \vspace{1.5mm}\\ 
\quad = \nabla^2 \rho({\bm r}) -  2\tau({\bm r}).
\end{array} \end{equation}
For the historical milestones for obtaining the Hamiltonian density in many-nucleon systems, see Refs.~\cite{engel-brink,vautherin}. \\

\subsubsection{Force depending on the parameter $t_0$}  
First, for the term with the coefficient $t_0$, we have
\[ \begin{array}{ll}
 (1 + x_0 P_{\sigma}) (1- P_r P_{\sigma} P_{\tau}) \vspace{1.5mm}\\
~=~  (1 + x_0 P_{\sigma}) (1  -  P_{\sigma} P_{\tau}) \vspace{1.5mm}\\
~=~  1 + x_0 P_{\sigma}  -  ( x_0 P_{\sigma}^2 + P_{\sigma})  P_{\tau} \vspace{1.5mm}\\ 
~=~  1 + \frac{1}{2} x_0 -  \frac{1}{2}  ( 1 + 2 x_0)  P_{\tau}  + \frac{1}{2} x_0 {\bm \sigma}_i  {\bm \sigma}_j   -  \frac{1}{2}  {\bm \sigma}_i {\bm \sigma}_j  P_{\tau}, 
 \end{array} \]
where we can take $P_r = 1$ due to the fact that the zero-range force acts only in S-waves.
It leads to the energy
\begin{equation} \label{t0ene}  \begin{array}{ll}
E_{t_0}  =  \frac{t_0}{2} \sum_{l,m} \vspace{1.5mm}\\
\qquad \int {\bar  \psi_l}({\bm r}_i') {\bar \psi_m}({\bm r}_j') 
 \left\{ 1 + \frac{1}{2} x_0 -  \frac{1}{2}  ( 1 + 2 x_0)  P_{\tau}  + \frac{1}{2} x_0 {\bm \sigma}_i  {\bm \sigma}_j   -  \frac{1}{2} {\bm \sigma}_i {\bm \sigma}_j  P_{\tau}
 \right\} \psi_l({\bm r}_i) \psi_m({\bm r}_j)  d{\bm r}_i d{\bm r}_j  d{\bm r}'_i d{\bm r}'_j |_{{\bm r}_i = {\bm r}_j = {\bm r}_i'= {\bm r}_j'} \vspace{1.5mm}\\
\quad = \int \large\{  \frac{t_0}{2}  (1 + \frac{1}{2} x_0 ) \rho({\bm r}_i,{\bm r}_i') \rho({\bm r}_i,{\bm r}_i') -  \frac{t_0}{2}(\frac{1}{2}+x_0)   \delta_{q_1,q_2}  \rho({\bm r}_i,{\bm r}_i') \rho({\bm r}_i,{\bm r}_i')  \vspace{1.5mm}\\
\qquad +  \frac{t_0}{4}x_0  {\bm s}({\bm r}_i,{\bm r}_i') {\bm s}({\bm r}_i,{\bm r}_i')  -  \frac{t_0}{4} \delta_{q_1,q_2}   {\bm s}({\bm r}_i,{\bm r}_i') {\bm s}({\bm r}_i,{\bm r}_i')  \large\} d{\bm r}_i d{\bm r}_j  d{\bm r}'_i d{\bm r}'_j |_{{\bm r}_i = {\bm r}_j = {\bm r}_i'= {\bm r}_j'}  \vspace{1.5mm}\\
\quad = \int dr^3 \left\{  \frac{t_0}{2}  (1 + \frac{1}{2} x_0 ) \rho({\bm r})^2 -  \frac{t_0}{2} (\frac{1}{2} + x_0) \sum_q \rho_q({\bm r})^2  +   \frac{t_0}{4}x_0  {\bm s}({\bm r})^2  -  \frac{t_0}{4} \sum_q  {\bm s}_q({\bm r})^2   \right\},
\end{array} \end{equation}
where $P_{\tau}$ is reduced to $\delta_{q_1,q_2}$ by assuming that there is no isospin mixing in nuclear force ($q_i$ identifies the isospin of the single particle state $i$).
The specific representation shown in the second line of Eq.~\eqref{t0ene}, which leads to the derivation of the Hamiltonian density, has been used in physics, where the property of $\delta$-function: $\int \delta({\bm a}-{\bm r}) f({\bm r}) d{\bm r} = f({\bm a})$ and therefore 
\[ \int \int \delta({\bm r}-{\bm r}') f({\bm r}') d{\bm r}' d{\bm r} = \int f({\bm r})d{\bm r} \]
is taken into account. \\

\subsubsection{Force depending on the parameter $t_1$}
Second, for the term with the coefficient $t_1$, we have
\[ \begin{array}{ll}  (1 + x_1 P_{\sigma}) ( {\bm k}^2 + {\bm k}'^2 ) (1- P_r P_{\sigma} P_{\tau})  \vspace{1.5mm}\\
\quad = \{ \left(\frac{1}{2i} (\nabla_i - \nabla_j) \right)^2 + \left( - \frac{1}{2i} (\nabla_i' - \nabla_j') \right)^2 \}  (1 + x_1 P_{\sigma}) (1- P_{\sigma} P_{\tau})  \vspace{1.5mm}\\
\quad  =  \{  -\frac{1}{4} (\nabla_i^2  - \nabla_i \nabla_j - \nabla_j \nabla_i  + \nabla_j ^2) -  \frac{1}{4} (\nabla_i'^2 - \nabla'_i \nabla_j' - \nabla_j' \nabla_i'  + \nabla_j'^2) \}  (1  - x_1 P_{\sigma}^2 P_{\tau}  + x_1 P_{\sigma} - P_{\sigma} P_{\tau})  \vspace{1.5mm}\\
\quad  = - \frac{1}{4}  (\nabla_i^2  + \nabla_j ^2   + \nabla_i'^2  + \nabla_j'^2  - 2  \nabla_i \nabla_j - 2 \nabla_i' \nabla_j') \{ 1- x_1 P_{\tau} +   \frac{1}{2} (x_1 - P_{\tau})(1 + {\bm \sigma}_i {\bm \sigma}_j)\}  \vspace{1.5mm}\\
\quad  = - \frac{1}{4}  (\nabla_i^2  + \nabla_j ^2   + \nabla_i'^2  + \nabla_j'^2  - 2  \nabla_i \nabla_j - 2 \nabla_i' \nabla_j') \{ 1  +   \frac{1}{2} x_1 - (x_1  +   \frac{1}{2}) P_{\tau}  +   \frac{1}{2} (x_1 - P_{\tau}){\bm \sigma}_i {\bm \sigma}_j \},
\end{array} \]
where we can take $P_r = 1$ due to the fact that the zero-range force acts only in S-waves.
It leads to the energy
\[ \begin{array}{ll}
E_{t_1} = - \frac{1}{2} \frac{t_1}{8} \sum_{l,m}  \int {\bar  \psi_l}({\bm r}_i') {\bar \psi_m}({\bm r}_j')
  (\nabla_i^2  + \nabla_j ^2   + \nabla_i'^2  + \nabla_j'^2  - 2  \nabla_i \nabla_j - 2 \nabla_i' \nabla_j')  \vspace{1.5mm}\\
\qquad \{ 1  +   \frac{1}{2} x_1 - (x_1  +   \frac{1}{2}) P_{\tau}  +   \frac{1}{2} (x_1 - P_{\tau}){\bm \sigma}_i {\bm \sigma}_j \} \psi_l({\bm r}_i) \psi_m({\bm r}_j)  d{\bm r}_i d{\bm r}_j  d{\bm r}'_i d{\bm r}'_j |_{{\bm r}_i = {\bm r}_j = {\bm r}_i'= {\bm r}_j'} \vspace{2.5mm}\\
\quad = - \frac{t_1}{16}  \int 
  (\nabla_i^2  + \nabla_j ^2   + \nabla_i'^2  + \nabla_j'^2  - 2  \nabla_i \nabla_j - 2 \nabla_i' \nabla_j')  \vspace{1.5mm}\\
\qquad \{ (1  +   \frac{1}{2} x_1)  \rho({\bm r}_i,{\bm r}_i') \rho({\bm r}_j,{\bm r}_j') - (x_1  +   \frac{1}{2})   \delta_{q_1,q_2}  \rho({\bm r}_i,{\bm r}_i') \rho({\bm r}_j,{\bm r}_j')  +   \frac{1}{2} x_1 {\bm s}({\bm r}_i,{\bm r}_i')  {\bm s}({\bm r}_j,{\bm r}_j')  \vspace{1.5mm}\\
\qquad - \frac{1}{2}   \delta_{q_1,q_2}   {\bm s}({\bm r}_i,{\bm r}_i') {\bm s}({\bm r}_j,{\bm r}_j')  \} d{\bm r}_i d{\bm r}_j  d{\bm r}'_i d{\bm r}'_j|_{{\bm r}_i = {\bm r}_j = {\bm r}_i'= {\bm r}_j'} \vspace{2.5mm}\\
\quad = - \frac{t_1}{16}  \int dr^3
 \{ ~ (1  +   \frac{1}{2} x_1) \{ [ (\nabla_i^2 + \nabla_i'^2) \rho({\bm r}_i,{\bm r}_i') \} \rho({\bm r}_j,{\bm r}_j')
+ \rho({\bm r}_i,{\bm r}_i')  (\nabla_j^2 + \nabla_j'^2)  \rho({\bm r}_j,{\bm r}_j')  \vspace{1.5mm}\\ 
\qquad - 2 \nabla_i \rho({\bm r}_i,{\bm r}_i')\nabla_j \rho({\bm r}_j,{\bm r}_j')
- 2 \nabla_i' \rho({\bm r}_i,{\bm r}_i')\nabla_j' \rho({\bm r}_j,{\bm r}_j')]  \vspace{1.5mm}\\
\qquad - (x_1  +   \frac{1}{2}) \sum_q   [ (\nabla_i^2 + \nabla_i'^2) \rho_q({\bm r}_i,{\bm r}_i') \} \rho_q({\bm r}_j,{\bm r}_j')
+ \rho_q({\bm r}_i,{\bm r}_i')  (\nabla_j^2 + \nabla_j'^2)  \rho_q({\bm r}_j,{\bm r}_j')  \vspace{1.5mm}\\ 
\qquad - 2 \nabla_i \rho_q({\bm r}_i,{\bm r}_i')\nabla_j \rho_q({\bm r}_j,{\bm r}_j')
- 2 \nabla_i' \rho_q({\bm r}_i,{\bm r}_i')\nabla_j' \rho_q({\bm r}_j,{\bm r}_j')]   \vspace{1.5mm}\\ 
\qquad  +   \frac{1}{2} x_1  [ (\nabla_i^2 + \nabla_i'^2) {\bm s}({\bm r}_i,{\bm r}_i') \}  {\bm s}({\bm r}_j,{\bm r}_j')
+  {\bm s}({\bm r}_i,{\bm r}_i')  (\nabla_j^2 + \nabla_j'^2)   {\bm s}({\bm r}_j,{\bm r}_j')  \vspace{1.5mm}\\ 
\qquad - 2 \nabla_i  {\bm s}({\bm r}_i,{\bm r}_i')\nabla_j  {\bm s}({\bm r}_j,{\bm r}_j')
- 2 \nabla_i'  {\bm s}({\bm r}_i,{\bm r}_i')\nabla_j'  {\bm s}({\bm r}_j,{\bm r}_j')]    \vspace{1.5mm}\\
\qquad - \frac{1}{2}  \sum_q   [ (\nabla_i^2 + \nabla_i'^2) {\bm s}_q({\bm r}_i,{\bm r}_i') \}  {\bm s}_q({\bm r}_j,{\bm r}_j')
+  {\bm s}_q({\bm r}_i,{\bm r}_i')  (\nabla_j^2 + \nabla_j'^2)  {\bm s}_q({\bm r}_j,{\bm r}_j') 
  \vspace{2.5mm}\\
\quad = - \frac{t_1}{16}  \int dr^3 
 \{ ~ (1  +   \frac{1}{2} x_1) [ 2( \nabla^2 \rho({\bm r}) -  2\tau({\bm r}) ) \rho({\bm r})
 -2 ( \frac{1}{2} \nabla  \rho({\bm r}) +  i{\bm j}({\bm r}))^2  -2 ( \frac{1}{2} \nabla  \rho({\bm r}) -  i{\bm j}({\bm r}))^2 ]  \vspace{1.5mm}\\
\qquad - (x_1  +   \frac{1}{2}) \sum_q  [ 2( \nabla^2 \rho_q({\bm r}) -  2\tau_q({\bm r}) ) \rho_q({\bm r})
 -2 ( \frac{1}{2} \nabla  \rho_q({\bm r}) +  i{\bm j}_q({\bm r}))^2  -2 ( \frac{1}{2} \nabla  \rho_q({\bm r}) -  i{\bm j}_q({\bm r}))^2 ]       \vspace{1.5mm}\\
\qquad +   \frac{1}{2} x_1  [ 2( \nabla^2 {\bm s}({\bm r}) -  2 {\bm T}({\bm r}) ) {\bm s}({\bm r})
 -2 ( \frac{1}{2} \nabla {\bm s}({\bm r}) +  i{\bm J}({\bm r}))^2  -2 ( \frac{1}{2} \nabla {\bm s}({\bm r}) -  i{\bm J}({\bm r}))^2 ]  \vspace{1.5mm}\\
\qquad - \frac{1}{2}  \sum_q  [ 2( \nabla^2 {\bm s}_q({\bm r}) -  2{\bm T}_q({\bm r}) )  {\bm s}_q({\bm r})
 -2 ( \frac{1}{2} \nabla   {\bm s}_q({\bm r}) +  i{\bm J}_q({\bm r}))^2  -2 ( \frac{1}{2} \nabla   {\bm s}_q({\bm r}) -  i{\bm J}_q({\bm r}))^2 ] \}   \vspace{2.5mm}\\
\quad = \int dr^3 \{  \frac{3 t_1}{16} [-(1 + \frac{1}{2}x_1)  \rho({\bm r}) \nabla^2  \rho({\bm r})+(x_1+ \frac{1}{2}) \sum_q  \rho_q({\bm r}) \nabla^2  \rho_q ({\bm r}) ] \vspace{1.5mm}\\
\qquad + \frac{t_1}{4} [ (1 + \frac{1}{2}x_1) (\rho({\bm r}) \tau({\bm r})-{\bm j}({\bm r})^2) -(x_1+ \frac{1}{2}) \sum_q  (\rho_q({\bm r}) \tau_q({\bm r})-{\bm j}_q({\bm r})^2) ] \vspace{1.5mm} \\
\qquad - \frac{3t_1}{32} [ x_1 {\bm s}({\bm r})  \nabla^2{\bm s}({\bm r}) -  \sum_q  {\bm s}_q({\bm r})  \nabla^2{\bm s}_q({\bm r}) ]  \vspace{1.5mm}\\
\qquad + \frac{t_1}{8} [x_1 ({\bm s}({\bm r}) \cdot{\bm T}({\bm r}) - {\bm J}^2({\bm r}) )    - \sum_q ({\bm s}_q({\bm r}) \cdot{\bm T}_q({\bm r}) - {\bm J}_q^2({\bm r}) )    ] \},
\end{array} \]
where we use the relations shown in Eq.~\eqref{tools} and the reduction: $P_{\tau} \to \delta_{q_1,q_2}$. \\

\subsubsection{Force depending on the parameter $t_2$}
Third, for the term with the coefficient $t_2$, we have
\[  \begin{array}{ll}
 (1+ x_2 P_\sigma )({\bm k} \cdot {\bm k}') (1 - P_{r} P_{\sigma}P_{\tau})  \vspace{1.5mm}\\
\quad = \{- \frac{1}{2i}(\nabla_i - \nabla_j) \times  \frac{-1}{2i}(\nabla_i' - \nabla_j')  \}  (1+ x_2 P_\sigma ) (1 + P_{\sigma}P_{\tau})   \vspace{1.5mm}\\
\quad =  \frac{1}{4}(\nabla_i \nabla_i' - \nabla_i \nabla_j' - \nabla_j \nabla_i' + \nabla_j \nabla_j' )   (1  + x_2 P_{\tau}  + x_2 P_\sigma  + P_{\sigma}P_{\tau})   \vspace{1.5mm}\\
\quad =  \frac{1}{4}(\nabla_i \nabla_i' - \nabla_i \nabla_j' - \nabla_j \nabla_i' + \nabla_j \nabla_j' )   (1  + x_2 P_{\tau}  + \frac{1}{2} (x_2  + P_{\tau})(1+{\bm \sigma}_i {\bm \sigma}_j) )   \vspace{1.5mm}\\
\quad =  \frac{1}{4}(\nabla_i \nabla_i' - \nabla_i \nabla_j' - \nabla_j \nabla_i' + \nabla_j \nabla_j' )   (1 + \frac{1}{2} x_2  + (x_2  + \frac{1}{2}) P_{\tau}  + \frac{1}{2} (x_2  + P_{\tau}){\bm \sigma}_i {\bm \sigma}_j ),  
\end{array}  \]
where we can take $P_r = -1$ since odd powers of ${\bm k}$ are treated.
It leads to the energy
\[ \begin{array}{ll}
E_{t_2} = \frac{1}{2}  \frac{t_2}{4} \sum_{l,m}  \int {\bar  \psi_l}({\bm r}_i') {\bar \psi_m}({\bm r}_j') (\nabla_i \nabla_i' - \nabla_i \nabla_j' - \nabla_j \nabla_i' + \nabla_j \nabla_j' )   \vspace{1.5mm} \\
(1 + \frac{1}{2} x_2  + (x_2  + \frac{1}{2}) P_{\tau}  + \frac{1}{2} (x_2  + P_{\tau}){\bm \sigma}_i {\bm \sigma}_j )  
  \psi_l({\bm r}_i)   \psi_m({\bm r}_j)  d{\bm r}_i d{\bm r}_j |_{{\bm r}_i = {\bm r}_j = {\bm r}_i'= {\bm r}_j'} \vspace{1.5mm}\\
 = \frac{t_2}{8}  \int  (\nabla_i \nabla_i' - \nabla_i \nabla_j' - \nabla_j \nabla_i' + \nabla_j \nabla_j' )  \vspace{1.5mm}  \\
\quad \{ (1 + \frac{1}{2} x_2)  \rho({\bm r}_i,{\bm r}_i') \rho({\bm r}_j,{\bm r}_j')
  +  (x_2  + \frac{1}{2}) \delta_{q_1,q_2}  \rho({\bm r}_i,{\bm r}_i') \rho({\bm r}_j,{\bm r}_j')  \vspace{1.5mm}  \\
\quad  + \frac{1}{2} x_2 {\bm s}({\bm r}_i,{\bm r}_i') {\bm s}({\bm r}_j,{\bm r}_j')   + \frac{1}{2}  \delta_{q_1,q_2} {\bm s}({\bm r}_i,{\bm r}_i') {\bm s}({\bm r}_j,{\bm r}_j') \} ~  d{\bm r}_i d{\bm r}_j |_{{\bm r}_i = {\bm r}_j = {\bm r}_i'= {\bm r}_j'} \vspace{1.5mm}\\
 = \frac{t_2}{4}   \int dr^3  [ (1 + \frac{1}{2} x_2) \{ \rho({\bm r})  \tau({\bm r})
- (\frac{1}{2} \nabla \rho({\bm r})  - {\bm j}({\bm r}))  (\frac{1}{2} \nabla \rho({\bm r})  + {\bm j}({\bm r}) ) \}  \vspace{1.5mm}  \\
\quad  +  (x_2  + \frac{1}{2}) \sum_q  \{ \rho_q({\bm r})  \tau_q({\bm r})
- (\frac{1}{2} \nabla \rho_q({\bm r})  - {\bm j}_q({\bm r}))  (\frac{1}{2} \nabla \rho_q({\bm r})  + {\bm j}_q({\bm r}) ) \}  \vspace{1.5mm}\\
\quad  + \frac{1}{2} x_2  \{ {\bm s}({\bm r}) \cdot {\bm T}({\bm r})
- (\frac{1}{2} \nabla  {\bm s}({\bm r})  - {\bm J}({\bm r}))  (\frac{1}{2} \nabla \cdot {\bm s}({\bm r})  + {\bm J}({\bm r})) \}  \vspace{1.5mm}\\
\quad  + \frac{1}{2} \sum_q  \{ {\bm s}_q({\bm r}) \cdot {\bm T}_q({\bm r})
- (\frac{1}{2} \nabla  {\bm s}_q({\bm r})  - {\bm J}_q({\bm r}))  (\frac{1}{2} \nabla \cdot {\bm s}_q({\bm r})  + {\bm J}_q({\bm r}) ) \}   ]   \vspace{1.5mm}\\
 = \frac{t_2}{16}   \int dr^3  [ (1 + \frac{1}{2} x_2) \{ 4 \rho({\bm r})  \tau({\bm r})
+   \rho({\bm r}) \nabla^2 \rho({\bm r})  - 4  {\bm j}({\bm r}) ^2 \}  \vspace{1.5mm}  \\
\quad  +  (x_2  + \frac{1}{2}) \sum_q  \{ 4 \rho_q({\bm r})  \tau_q({\bm r})
+   \rho_q({\bm r}) \nabla^2 \rho({\bm r})  - 4  {\bm j}_q({\bm r}) ^2 \}  \vspace{1.5mm}\\
\quad  +  x_2  \{ 2 {\bm s}({\bm r}) \cdot {\bm T}({\bm r})
+ \frac{1}{2} {\bm s}({\bm r}) \cdot \nabla^2 {\bm s}({\bm r})  - 2 {\bm J}({\bm r}) ^2 \} \vspace{1.5mm}\\
\quad  + \sum_q  \{ 2 {\bm s}_q({\bm r}) \cdot {\bm T}_q({\bm r})
+ \frac{1}{2}  {\bm s}_q({\bm r}) \cdot \nabla^2 {\bm s}_q({\bm r})  - 2 {\bm J}({\bm r}) ^2 \} ],  
\end{array} \]
where we use the relations shown in Eq.~\eqref{tools} and the reduction: $P_{\tau} \to \delta_{q_1,q_2}$. \\

\subsubsection{Force depending on the parameter $t_3$}
Fourth, for the term with the coefficient $t_3$ (the density-dependent term), we have
\[  \begin{array}{ll}
 (1+ x_3 P_\sigma )  \rho(\frac{{\bm r}_i + {\bm r}_j}{2})^{\alpha} (1 - P_{r} P_{\sigma}P_{\tau})  \vspace{1.5mm}\\
\quad =  (1+ x_3 P_\sigma )   (1 - P_{\sigma}P_{\tau})  \rho (\frac{{\bm r}_i + {\bm r}_j}{2})^{\alpha}   \vspace{1.5mm}\\
\quad =  \{ 1+ x_3 P_\sigma    -   P_{\sigma}P_{\tau}  - x_3  P_{\sigma}^2  P_{\tau} \}  \rho(\frac{{\bm r}_i + {\bm r}_j}{2})^{\alpha}   \vspace{1.5mm}\\   
\quad = \{ 1 + \frac{1}{2} x_3 -  \frac{1}{2}  ( 1 + 2 x_3)  P_{\tau}  + \frac{1}{2} x_3 {\bm \sigma}_i  {\bm \sigma}_j   -  \frac{1}{2}  {\bm \sigma}_i {\bm \sigma}_j  P_{\tau} \}  \rho(\frac{{\bm r}_i + {\bm r}_j}{2})^{\alpha}, 
\end{array}  \]
where we can take $P_r = 1$ due to the fact that the zero-range force acts only in S-waves.
It leads to the energy
\[ \begin{array}{ll}
E_{t_3} = \frac{1}{2}  \frac{t_3}{6} \sum_{l,m}  \int {\bar  \psi_l}({\bm r}_i') {\bar \psi_m}({\bm r}_j') \left\{ 
1 + \frac{1}{2} x_3 -  \frac{1}{2}  ( 1 + 2 x_3)  P_{\tau}  + \frac{1}{2} x_3 {\bm \sigma}_i  {\bm \sigma}_j   -  \frac{1}{2} {\bm \sigma}_i {\bm \sigma}_j  P_{\tau}
 \right\}  \vspace{1.5mm}\\
\qquad  \rho(\frac{{\bm r}_i + {\bm r}_j}{2})^{\alpha}   \psi_l({\bm r}_i)   \psi_m({\bm r}_j)  d{\bm r}_i d{\bm r}_j |_{{\bm r}_i = {\bm r}_j = {\bm r}_i'= {\bm r}_j'} \vspace{1.5mm}\\
\quad = \int \large\{  \frac{t_3}{12}  (1 + \frac{1}{2} x_3 ) \rho({\bm r}_i,{\bm r}_i') \rho({\bm r}_i,{\bm r}_i') -  \frac{t_3}{12}(\frac{1}{2}+x_3)   \delta_{q_1,q_2}  \rho({\bm r}_i,{\bm r}_i') \rho({\bm r}_i,{\bm r}_i')  \vspace{1.5mm}\\
\qquad +  \frac{t_3}{24}x_3  {\bm s}({\bm r}_i,{\bm r}_i') {\bm s}({\bm r}_i,{\bm r}_i')  -  \frac{t_3}{24} \delta_{q_1,q_2}   {\bm s}({\bm r}_i,{\bm r}_i') {\bm s}({\bm r}_i,{\bm r}_i')   \rho(\frac{{\bm r}_i + {\bm r}_j}{2})^{\alpha}   \large\} d{\bm r}_i d{\bm r}_j |_{{\bm r}_i = {\bm r}_j = {\bm r}_i'= {\bm r}_j'}  \vspace{1.5mm}\\
\quad = \int \left\{  \frac{t_3}{12}  (1 + \frac{1}{2} x_3 ) \rho({\bm r})^{2+\alpha} -  \frac{t_3}{12} (\frac{1}{2} + x_3) \sum_q \rho_q({\bm r})^2   \rho_q({\bm r})^{\alpha} +   \frac{t_3}{24}x_3  {\bm s}({\bm r})^2  \rho({\bm r})^{\alpha}  -  \frac{t_3}{24} \sum_q  {\bm s}_q({\bm r})^2  \rho_q({\bm r})^{\alpha}   \right\}    d{\bm r}^3,
\end{array} \]
where we use the reduction: $P_{\tau} \to \delta_{q_1,q_2}$. \\

\subsubsection{Hamiltonian density of interacting many-nucleon systems}
The expectation value $E$ of the many-body Hamiltonian is written by an integral of the Hamiltonian density $H({\bm r})$.
\[ E =  E_{t_0} +  E_{t_1} +  E_{t_2} +  E_{t_3}  = \int dr^3 H({\bm r}). \]
The Hamiltonian density is obtained by collecting the results:
\begin{equation} \begin{array}{ll} \label{hamdens} 
H({\bm r}) = \frac{1}{2} \big[ \frac{\hbar^2}{2m}{\tau} + \frac{t_0}{2}  (1 + \frac{1}{2} x_0 ) \rho^2 -  \frac{t_0}{2} (\frac{1}{2} + x_0) \sum_q \rho_q^2  +   \frac{t_0}{4}x_0  {\bm s}^2  -  \frac{t_0}{4} \sum_q  {\bm s}_q^2  \vspace{2.5mm} \\
\quad -\frac{3 t_1}{16}(1 + \frac{1}{2}x_1) \rho \triangle  \rho +  \frac{3 t_1}{16} (x_1+ \frac{1}{2}) \sum_q  \rho_q \triangle  \rho_q  + \frac{t_1}{4}  (1 + \frac{1}{2}x_1) (\rho \tau -{\bm j}^2) - \frac{t_1}{4} (x_1+ \frac{1}{2}) \sum_q  (\rho_q \tau_q -{\bm j}_q^2) \vspace{1.5mm} \\
\quad - \frac{3t_1}{32}  x_1 {\bm s}  \triangle {\bm s} +  \frac{3t_1}{32}  \sum_q  {\bm s}_q \triangle {\bm s}_q  + \frac{t_1}{8} x_1 ({\bm s} \cdot{\bm T} - {\bm J}^2 )    -  \frac{t_1}{8} \sum_q ({\bm s}_q \cdot{\bm T}_q - {\bm J}_q^2 )   ]  \vspace{2.5mm}\\
 \quad + \frac{t_2}{16}  (1 + \frac{1}{2} x_2)  \rho \triangle \rho + \frac{t_2}{4}  (1 + \frac{1}{2} x_2) ( \rho \tau  - {\bm j}^2 )
+  \sum_q  \{  \frac{t_2}{16} (x_2  + \frac{1}{2})  \rho_q \triangle \rho +  \frac{t_2}{4}  (x_2  + \frac{1}{2})( \rho_q  \tau_q - {\bm j}_q^2)   \}  \vspace{1.5mm}\\
\quad +  \frac{t_2}{32}  x_2 {\bm s} \cdot \triangle {\bm s} +  \frac{t_2}{8}  x_2 ({\bm s} \cdot {\bm T} -  {\bm J} ^2 ) 
 + \sum_q  \{  \frac{t_2}{32} {\bm s}_q \cdot \triangle {\bm s}_q +  \frac{t_2}{8}  ({\bm s}_q \cdot {\bm T}_q-  {\bm J}_q^2 ) \}  \vspace{2.5mm}\\  
\quad  + \frac{t_3}{12}  (1 + \frac{1}{2} x_3 ) \rho^{2+\alpha} -
  \frac{t_3}{12} (\frac{1}{2} + x_3) \sum_q \rho_q^2   \rho_q^{\alpha} +   \frac{t_3}{24}x_3  {\bm s}^2  \rho^{\alpha}  -  \frac{t_3}{24} \sum_q  {\bm s}_q^2  \rho^{\alpha} \big] ;  
\end{array} \end{equation}
it is half of the standard Hamiltonian density~\cite{engel-brink,vautherin}, which can be understood by applying the variational principle to the first term of the right hand side (the contribution from the kinetic energy), where $1/2$ for the interaction part appears to avoid calculating the interaction twice.
The justification of $1/2$ (in the term arising from the kinetic energy) is illustrated in the next section (Sec.~\ref{sub3}).
It is notable that differential operators appear due to the non-local interaction.
Interacting many-nucleon systems cannot be described well if differential operators are not utilized (cf. local density approximation). 
 \vspace{16mm} \\

\subsection{Energy density functional  \label{sub3}}

The components in the effective interaction are obtained based on the variational principle.
This corresponds to the procedure of obtaining the functional representation of an effective Hamiltonian.
The following points, which are illustrated in concrete discussion, should be noticed:
\begin{itemize}
\item Once the energy is given, it is not necessary to have the fermionic relation or the Slater-determinant formalism to derive the effective interaction (the effective Hamiltonian).
\item The boundary condition is indispensable. 
\item Regularity of wave functions is required; i.e., $L^2$-space is not necessarily sufficient. 
\
\item For mathematically strict treatment, the variational principle should be considered not only in the real Hilbert/Banach spaces (by varying ``real'' densities) but in the full complex Hilbert/Banach spaces (by varying ``complex'' wave functions).
\end{itemize}
Since the application of the variational principle includes differentiation, more careful treatment is required (differentiability and so on).
Furthermore, the application of variational principle means a process of obtaining a functional representation in some complex functional spaces only from the real function (i.e., energy). 
In order to present the essential treatment of the variational principle, the discussion is developed for a spin-saturated and charge-conjugate nucleus (i.e., $\rho/2 = \rho_p = \rho_n$).
In this situation Eq.~\eqref{hamdens} reduces to a model Hamiltonian density:
\begin{equation} \begin{array}{ll} \label{hamdensexam}
{\mathcal H}({\bm r}) = {\tilde t}_1{\tau} + {\tilde t}_2 \rho^2  + {\tilde t}_3 \rho^{\tilde \alpha}  + {\tilde t}_4 (\rho \tau -{\bm j}^2)  +{\tilde t}_5  \rho \triangle  \rho,    
\end{array} \end{equation}
where ${\tilde \alpha}$ is a positive rational number, and $\{ {\tilde t}_i \}$ is a parameter set related to the parameter sets $\{ t_i \}$ and $\{ x_i \}$~\cite{engel-brink}.
Here, based on this model Hamiltonian, generalized higher order terms and terms with fractional powers are also treated, where we need to discuss within suitable functional spaces for a strict treatment of the variational principle. 
First, the term arising from the kinetic energy (the homogeneous term of Schr\"odinger equations) is treated in Ex. \ref{ex0}.
Next, the derivation of terms arising from the interaction energy is demonstrated.

\subsubsection{Linear and nonlinear Laplacians} \label{ex0}
Let $L^p (\Omega)$ be the space of functions on $\Omega$ which are $L^p$ for the Lebesgue measure, where $L^p (\Omega)$ are not (infinite-dimensional) Hilbert spaces but (infinite-dimensional) Banach spaces if $p$ is not equal to 2.
Let $p$ be an even number satisfying $p \ge 2$ and $\psi$ be included in $W^{1,p}(\Omega)$ with a open bounded set $\Omega \subset {\mathbb R}^3$, where $W^{1,p}(\Omega)$ is the space of functions in $L^p(\Omega)$ whose distribution derivatives of order $\le 1$ are in $L^p(\Omega)$ (for the ``mathematical'' distribution~\cite{schwartz}, refer to the textbook of functional analysis).
For our purposes it is necessary to introduce such a generalized differential in association with the spaces of integrable functions. 
Note here that the $\delta$-function, which cannot be defined using usual functions ($C^k$-function and so on), is successfully defined in the sense of a distribution.
In this sense the zero-range interaction formalism is based on the theory of distribution.
 
First of all it is useful to consider a simple case when the energy is represented by
\[{\mathcal E}(\psi) = \frac{1}{p} \int_{\Omega} dr^3  |\nabla \psi |^p,  \]
where the corresponding Hamiltonian energy density is ${\mathcal H}(\psi) = \frac{1}{p} | \nabla \psi |^p$.
The differential operator is considered in the sense of distribution; this is true to other cases shown in this section (Sec~\ref{sub3}). 
By minimizing this energy (i.e., the Gateaux differential: ${\mathcal E}'(\psi) = {\mathcal A}$ is considered), we obtain
\[{\mathcal A}(\psi) = - \nabla \cdot (|\nabla \psi|^{p-2} \nabla \psi). \]
Indeed, for $\psi$ and $\phi$ in $W^{1,p}(\Omega)$,
\begin{equation} \label{deriv} \begin{array}{ll}
|\nabla \psi + \lambda \phi|^p - |\nabla \psi|^p = \int_0^1 \frac{\partial}{\partial t} |\nabla \psi + t \lambda \nabla \phi|^p dt  \vspace{1.5mm}  \\
\qquad = \int_0^1 \frac{\partial}{\partial t} (|\nabla \psi + t \lambda \nabla \phi|^2)^{p/2} dt  \vspace{1.5mm}  \\
\qquad = \int_0^1 \frac{p}{2} (|\nabla \psi + t \lambda \nabla \phi|^2)^{(p-2)/2}  \frac{\partial}{\partial t} |\nabla \psi + t \lambda \nabla \phi|^2 dt   \vspace{1.5mm}  \\
\qquad = \int_0^1 \frac{p}{2} (|\nabla \psi + t \lambda \nabla \phi|^2)^{(p-2)/2}  \cdot 2 {\rm Re} \{ (\nabla \psi + t \lambda \nabla \phi) \cdot \lambda \nabla {\bar \phi} \} dt \vspace{1.5mm}  \\
\qquad =  p \lambda \int_0^1 |\nabla \psi + t \lambda \nabla \phi|^{p-2}  \cdot {\rm Re} \{ (\nabla \psi + t \lambda \nabla \phi) \cdot \nabla {\bar \phi} \} dt  
\end{array} \end{equation}
is true.
It follows that
\[ \begin{array}{ll}
 \frac{{\mathcal E}(\psi + \lambda \phi) - {\mathcal E}(\psi)}{\lambda}
= \int_{\Omega} dr^3 ~  \int_0^1 |\nabla \psi + t \lambda \nabla \phi|^{p-2}  {\rm Re} \{ (\nabla \psi + t \lambda \nabla \phi) \cdot \nabla {\bar \phi} \} dt.
\end{array} \]
Due to $\lambda \to 0$ ($\lambda > 0$),
\[ |\nabla \psi + t \lambda \nabla \phi|^{p-2}  {\rm Re} \{ (\nabla \psi + t \lambda \nabla \phi) \cdot \nabla {\bar \phi} \}
 ~\to~ |\nabla \psi|^{p-2}   {\rm Re} (\nabla \psi \cdot \nabla {\bar \phi}) \quad {\rm a.e.}  \] 
is valid (``a.e.'' means ``almost everywhere'', whose mathematical definition should be referred to the textbook of functional analysis.
Furthermore, due to $\lambda \to 0$, the following inequality is valid:
\[ \left| |\nabla \psi + t \lambda \nabla \phi|^{p-2}  {\rm Re} \{ (\nabla \psi + t \lambda \nabla \phi) \cdot \nabla {\bar \phi}  \} \right|
~\le~  |\nabla \psi + t \lambda \nabla \phi|^{p-1}  |\nabla \phi| 
~\le~ ( |\nabla \psi| + |\lambda| |\nabla \phi|)^{p-1}  |\nabla \phi|,  \]
where it is worth noting here that the last term is integrable.
We have
\[ \begin{array}{ll}
\left. \frac{d}{d \lambda} {\mathcal E} (\psi + \lambda \phi) \right|_{\lambda = 0}
= \lim_{\lambda \to 0}  \frac{{\mathcal E}(\psi + \lambda \phi) - {\mathcal E}(\psi)}{\lambda}
= \int_{\Omega} dr^3 ~  |\nabla \psi|^{p-2}   {\rm Re} (\nabla \psi \cdot \nabla {\bar \phi})
= {\rm Re} \int_{\Omega} dr^3 ~  |\nabla \psi|^{p-2}   \nabla \psi \cdot \nabla {\bar \phi}.
\end{array} \]
If $\psi$ and $\phi$ are sufficiently smooth,
\[ \int_{\Omega}  |\nabla \psi|^{p-2}   \nabla \psi \cdot \nabla {\bar \phi}  ~dr^3  =  \int_{\partial \Omega}  |\nabla \psi|^{p-2} \frac{\partial \psi}{\partial \mu} {\bar \phi} ~dS + \int_{\Omega}  {\mathcal A}(\psi) {\bar \phi} ~dr^3  \]
follows from the integration by parts, where $\partial/\partial \mu$ means the outward normal differential operator.
Although we have discussed the limit-process in a roundabout sort of way, it is based on the fact that $\lim_{\lambda \to 0} \frac{{\mathcal E}(\psi + \lambda \phi) - {\mathcal E}(\psi)}{\lambda} = - \int_{\Omega} dr^3 | \nabla \psi|^{p-2} \nabla \psi \nabla {\bar \phi}$ cannot follow from some simple treatments, because the left hand side is real-valued and the the right hand side is complex-valued.
It means that the differentiability of the energy cannot follow naively (for the differentiability, see Appendix (Sec.~\ref{appdx})). 
This fact is true for all the cases treated in Sec.~\ref{sub3}.

The final treatment actually depends on the boundary condition. 
When the Dirichlet boundary condition is adopted (for the homogeneous part, Schr\"odinger operator is considered), $W^{1,p}(\Omega)$ is replaced by $W^{1,p}_0(\Omega)$, and we have 
\begin{equation} \label{linearop} \int_{\Omega} ~  |\nabla \psi|^{p-2}   \nabla \psi \cdot \nabla {\bar \phi} ~dr^3  = \int_{\Omega}  {\mathcal A}(\psi) {\bar \phi} ~dr^3  = ({\mathcal A}(\psi), \phi).  \end{equation} 
That is, for $f \in W^{-1,p'}(\Omega) =  W^{1,p}_0(\Omega)^*$ and $\psi \in W_0^{1,p}(\Omega)$, the equality: $\int_{\Omega} ~  |\nabla \psi|^{p-2}   \nabla \psi \cdot \nabla {\bar \phi} ~dr^3  = (f,\phi)$ is valid to any $\psi \in W_0^{1,p}(\Omega)$, which is an generalized solution of 
\[ \left\{ \begin{array}{ll}
{\mathcal A}(\psi) = f \quad {\rm in} ~ \Omega,  \vspace{1.5mm} \\
\psi = 0 \quad {\rm on} ~  \partial \Omega.
\end{array} \right. \]
Meanwhile when the Neumann boundary condition is adopted, we have Eq.~\eqref{linearop}.
However the detail is different from the case with the Dirichlet boundary condition; for $f \in W^{1,p}(\Omega)^*$, the equality: $\int_{\Omega} ~  |\nabla \psi|^{p-2}   \nabla \psi \cdot \nabla {\bar \phi} ~dr^3  = (f,\phi)$ is valid to any $\psi \in W^{1,p}(\Omega)$, which is an generalized solution of 
\[ \left\{ \begin{array}{ll}
{\mathcal A}(\psi) = f \quad {\rm in} ~ \Omega,  \vspace{1.5mm} \\
\frac{\partial \psi}{\partial \mu} = 0 \quad {\rm on} ~  \partial \Omega. 
\end{array} \right. \]

When the periodic boundary condition is adopted for cuboid $\Omega$, $W^{1,p}(\Omega)$ is replaced by $W^{1,p}_{per}(\Omega)$, which is the space of restrictions to $\Omega$ of periodic functions. 
We have Eq.~\eqref{linearop} using the similar argument.
As a result, 
\begin{equation} \begin{array}{ll}
\left. \frac{d}{d \lambda} {\mathcal E} (\psi + \lambda \phi) \right|_{\lambda = 0}
= {\rm Re} \int_{\Omega} dr^3 ~  |\nabla \psi|^{p-2}   \nabla \psi \cdot \nabla {\bar \phi}
 ={\rm Re} \int_{\Omega}  {\mathcal A}(\psi) {\bar \phi} ~dr^3  =  {\rm Re} ({\mathcal A}(\psi), \phi) 
\end{array} \end{equation}
is valid to the Dirichlet, Neumann and periodic boundary conditions.
The operator ${\mathcal A}$ with $p>2$ is called $p-$Laplacian or nonlinear Laplacian.
If we take $p =2$, the relation between ${\mathcal A}$, ${\mathcal H}$ and ${\mathcal E}$ is that between the Laplacian, the corresponding Hamiltonian density and the corresponding energy.
The case with $p=2$ has the direct connection to the standard kinetic energy term.
Note that, for this discussion, it is sufficient that $\psi$ is included in $W^{1,p}(\Omega)$, and it is not necessary for $\psi$ to be a component of the Slater determinant. 

In case of many-particle situations there is no need to introduce extraordinary treatment.
Let the many-particle wave function be
\[ \Psi = \sum_{l=1}^N \psi_l.  \]
It is trivial that the energy becomes
\[{\mathcal E}_N(\Psi) = \frac{1}{p} \int_{\Omega} dr^3  \sum_{l=1}^N  |\nabla_l \psi_l |^p,  \]
where $\nabla_i$ acts on $\psi_i$, and the term in the effective Hamiltonian
\[{\mathcal A}_N(\Psi) = \sum_{l=1}^N  - \nabla_l \cdot (|\nabla_l \psi_l|^{p-2} \nabla_l \psi_l) \]
follows. 
Here the procedure of obtaining a term in the effective Hamiltonian from a given energy consists only of Gateaux differential, so that the similar procedure is also valid if we assume many-particle situations.
When we consider the linear situation $p=2$, the term is equal to $\sum_{i=1}^N ( - \triangle_i \psi_i)$, corresponding to the homogeneous term of the typical type of many-body Schr\"odinger equations.
In the following we do not discuss the many-particle situations explicitly, but ,as is the previously shown, they trivially follow once the corresponding one-particle situations are well understood.

\subsubsection{Nonlinear interactions depending on the parameters ${\tilde t}_2$ and ${\tilde t}_3$} \label{ex1}
Again let $p \in {\mathbb Z}$ ($\mathbb Z$: a set of all integers), be an even number satisfying $p>2$.
We consider the energy represented by
\begin{equation} \label{enerho}
{\mathcal E}_{D,Z}(\psi) = \frac{1}{p} \int_{\Omega} dr^3  | \psi |^{p},  
\end{equation}
where the corresponding Hamiltonian energy density is ${\mathcal H}_{D,Z} (\psi) = \frac{1}{p} | \psi |^p$. 
By minimizing this energy (i.e., the Gateaux differential is considered), we obtain
\[ \begin{array}{ll}
{\mathcal F}_{D,Z}(\psi) = |\psi|^{p-2} \psi,
\end{array} \]
where this is one of the most typical ingredient of nonlinear Schr\"odinger equations.
Indeed, for $\psi$ and $\phi$ in $L^p(\Omega)$,
\begin{equation} \label{deriv3} \begin{array}{ll}
|\psi + \lambda \phi|^p - | \psi|^p = \int_0^1 \frac{\partial}{\partial t} | \psi + t \lambda \phi|^p dt
=  p \lambda \int_0^1 |\psi + t \lambda \phi|^{p-2}  \cdot {\rm Re} \{ (\psi + t \lambda \phi) \cdot {\bar \phi} \} dt  
\end{array} \end{equation}
is true.
\begin{equation} \label{deriv4}
|\psi + t \lambda \phi|^{p-2}  {\rm Re} \{ (\psi + t \lambda \phi) \cdot {\bar \phi} \}
 ~\to~ |\psi|^{p-2}   {\rm Re} (\psi \cdot {\bar \phi}) \quad {\rm a.e.}  \end{equation}
is valid due to $\lambda \to 0$ ($\lambda > 0$).
In the same manner, we have
\[ \begin{array}{ll}
\left. \frac{d}{d \lambda} {\mathcal E}_{D,Z} (\psi + \lambda \phi) \right|_{\lambda = 0}
= \lim_{\lambda \to 0}  \frac{{\mathcal E}_{D,Z}(\psi + \lambda \phi) - {\mathcal E}_{D,Z}(\psi)}{\lambda}
= \int_{\Omega} dr^3 ~  |\psi|^{p-2}   {\rm Re} (\psi \cdot {\bar \phi})
= {\rm Re} \int_{\Omega} dr^3 ~  |\psi|^{p-2}  \psi \cdot {\bar \phi}.
\end{array} \]
As a result, we have
\begin{equation} \begin{array}{ll}
\left. \frac{d}{d \lambda} {\mathcal E}_{D,Z} (\psi + \lambda \phi) \right|_{\lambda = 0}
= {\rm Re} \int_{\Omega} dr^3 ~  | \psi|^{p-2}  \psi \cdot {\bar \phi}
= {\rm Re} \int_{\Omega}  {\mathcal F}_{D,Z}(\psi) {\bar \phi} ~dr^3  =  {\rm Re} ({\mathcal F}_{D,Z}(\psi), \phi). 
\end{array} \end{equation}
Nonlinear Schr\"odinger equations only including the interaction ${\mathcal F}_{D,Z}(\psi)$ with $p=4$ are known as one of the most typical type of nonlinear Schr\"odinger equations (cf. Ginzburg-Landau formalism). 

Let us consider the energy ${\mathcal E}_{D,Q}$ when $p \in {\mathbb Z}$ in Eq.~\eqref{enerho} is replaced by $q \in {\mathbb Q}$ ($\mathbb Q$: a set of all rational numbers). 
Nonlinear interaction with $q \in {\mathbb Q}$, which is simply called density-dependent force, is suggested to be indispensable to explain experimental results:
\[ \begin{array}{ll}
{\mathcal F}_{D,Q}(\psi) = |\psi|^{q-2} \psi,
\end{array} \]
where $q$ satisfies $q>2$.
For example, $q$ satisfying ``$(q-2)/2 = 1+1/6$'' has been proposed as a possible candidate~\cite{chabanat}, as well as that satisfying $(q-2)/2 = 1+1/4$~\cite{reinhard}.
Note here that the ``density-dependent'' force is a technical term for the force ${\mathcal F}_{D,Q}(\psi)$ satisfying $q \ne 4$. 
If we simply use the previous discussion shown in Eqs.~\eqref{deriv3} and \eqref{deriv4}, it is necessary to introduce the $L^{q}$-space of the fractional power.
Here is not problem to define $L^{q}$-space with noninteger $q$, and such $L^{q}$-space satisfying $1 \le q \le \infty$ holds the property of Banach spaces.
Accordingly the discussion shown in Eqs.~\eqref{deriv3} and \eqref{deriv4} are valid even in this case, and we have
\begin{equation} \begin{array}{ll}
\left. \frac{d}{d \lambda} {\mathcal E}_{D,Q} (\psi + \lambda \phi) \right|_{\lambda = 0}
= {\rm Re} \int_{\Omega} dr^3 ~  | \psi|^{q-2}  \psi \cdot {\bar \phi}
= {\rm Re} \int_{\Omega}  {\mathcal F}_{D,Q}(\psi) {\bar \phi} ~dr^3  =  {\rm Re} ({\mathcal F}_{D,Q}(\psi), \phi).
\end{array} \end{equation}

\subsubsection{Nonlinear interaction depending on the parameter ${\tilde t}_4$} \label{ex2}
Again, let $p \in {\mathbb Z}$ ($\mathbb Z$: a set of all integers), be an even number satisfying $p>2$.
We consider the energy represented by
\[{\mathcal E}_{K,Z}(\psi) = \frac{1}{p} \int_{\Omega} dr^3  |\nabla  \psi|^{p-2} |\psi|^2,  \]
where the corresponding Hamiltonian energy density is ${\mathcal H}_{K,Z} (\psi) = \frac{1}{p} |\nabla  \psi|^{p-2} |\psi|^2 $. 
By minimizing this energy (i.e., the Gateaux differential is considered), we obtain
\[ \begin{array}{ll}
{\mathcal F}_{K,Z}(\psi) = ~ - \frac{p-2}{p} \nabla \cdot ( |\nabla \psi|^{p-4} |\psi|^2  \nabla \psi) + \frac{2}{p}  |\nabla \psi|^{p-2} \psi.
\end{array} \]
Indeed, for $\psi$ and $\phi$ in $W^{1,p}(\Omega)$,
\begin{equation} \label{deriv} \begin{array}{ll}
|\nabla (\psi + \lambda \phi)|^{p-2}|\psi + \lambda \phi|^2  - |\nabla \psi|^{p-2} |\psi|^2 
 = \int_0^1 \frac{\partial}{\partial t} \{ |\nabla (\psi + t \lambda \phi)|^{p-2}  |\psi + t \lambda \phi|^2 \}~  dt \vspace{1.5mm}\\
\quad = \lambda  \int_0^1 \big[ (p-2) |\nabla (\psi + t \lambda \phi)|^{p-4}  |\psi + t \lambda \phi|^2  \cdot {\rm Re} \{ \nabla (\psi + t \lambda \phi) \cdot \nabla {\bar \phi} \}  \vspace{1.5mm}\\
\qquad + 2 |\nabla (\psi + t \lambda \phi)|^{p-2}  \cdot {\rm Re} \{(\psi + t \lambda \phi) \cdot {\bar \phi} \} \big]  dt
\end{array} \end{equation}
is true.
\begin{equation} \label{deriv2}  \begin{array}{ll}
(p-2) |\nabla (\psi + t \lambda \phi)|^{p-4}  |\psi + t \lambda \phi|^2  \cdot {\rm Re} \{ \nabla (\psi + t \lambda \phi) \cdot \nabla {\bar \phi} \}  ~\to~  (p-2) |\nabla \psi|^{p-4}  |\psi|^2  {\rm Re} ( \nabla \psi \cdot \nabla {\bar \phi} )  \quad {\rm a.e.},  \vspace{1.5mm}\\
 2 |\nabla (\psi + t \lambda \phi)|^{p-2}  \cdot {\rm Re} \{(\psi + t \lambda \phi) \cdot {\bar \phi} \}
 ~\to~  2 |\nabla \psi|^{p-2}   {\rm Re} ( \psi \cdot {\bar \phi} ) \quad {\rm a.e.}
\end{array}  \end{equation}
are valid due to $\lambda \to 0$ ($\lambda > 0$).
We have
\[ \begin{array}{ll}
\left. \frac{d}{d \lambda} {\mathcal E}_{K,Z} (\psi + \lambda \phi) \right|_{\lambda = 0}
= {\rm Re} \frac{1}{p}  \int_{\Omega} dr^3 ~ \{  (p-2) |\nabla \psi|^{p-4}  |\psi|^2  ( \nabla \psi \cdot \nabla {\bar \phi})  +  2 |\nabla \psi|^{p-2}  (\psi \cdot {\bar \phi}) \}.
\end{array} \]
If $\psi$ and $\phi$ are sufficiently smooth,
\[ \int_{\Omega} |\nabla \psi|^{p-4} |\psi|^2  \nabla \psi \cdot \nabla {\bar \phi}  ~dr^3  =  \int_{\partial \Omega}  |\nabla \psi|^{p-4}  |\psi|^2 \frac{\partial \psi}{\partial \mu} {\bar \phi} ~dS 
 - {\rm Re} \int_{\Omega} \nabla \cdot ( |\nabla \psi|^{p-4} |\psi|^2  \nabla \psi) \cdot {\bar \phi}  ~dr^3.  \]
As a result, by taking into account the boundary condition: the boundary conditions shown in Sec.~\ref{ex0}, we have
\begin{equation} \begin{array}{ll}
\left. \frac{d}{d \lambda} {\mathcal E}_{K,Z} (\psi + \lambda \phi) \right|_{\lambda = 0}
=  - {\rm Re}   \int_{\Omega} dr^3 ~ \frac{p-2}{p} \nabla \cdot ( |\nabla \psi|^{p-4} |\psi|^2  \nabla \psi) \cdot {\bar \phi}
+ {\rm Re}   \int_{\Omega} dr^3 ~  \frac{2}{p}  |\nabla \psi|^{p-2} \psi \cdot {\bar \phi}
\vspace{1.5mm} \\
= {\rm Re} \int_{\Omega}  {\mathcal F}_{K,Z}(\psi) {\bar \phi} ~dr^3  =  {\rm Re} ({\mathcal F}_{K,Z}(\psi), \phi).
\end{array} \end{equation}
Note that this interaction including differential operators cannot be bounded on $L^2(\Omega)$.
If $p=4$, ${\mathcal F}_{K,Z}$ is reduced to
\[ \begin{array}{ll}
{\mathcal F}_{K,Z}(\psi) = ~ - \frac{1}{2} \nabla \cdot ( |\psi|^2  \nabla \psi)  + \frac{1}{2}  |\nabla \psi|^{2} \psi.
\end{array} \]

Similar to Sec.~\ref{ex1}, nonlinear interaction with $q \in {\mathbb Q}$ is derived from the energy $E_{K,Q}$, which is obtained by replacing $p \in {\mathbb Z}$ in $E_{K,Z}$ by $q \in {\mathbb Q}$.
\[ \begin{array}{ll}
{\mathcal F}_{K,Q}(\psi)  = ~ - \frac{q-2}{q} \nabla \cdot ( |\nabla \psi|^{q-4} |\psi|^2  \nabla \psi) + \frac{2}{q}  |\nabla \psi|^{q-2} \psi.
\end{array} \]
where $q$ satisfies $q>2$.
If we simply use the previous discussion, it is necessary to introduce the $W^{1,q}$-space of the fractional power.
Using the interpolation of Sobolev spaces, there is no problem to define $W^{1,q}$-space with noninteger $q$, and such $W^{1,q}$-space satisfying $1 \le q \le \infty$ holds the property of Banach spaces.
Accordingly the discussion shown in Eqs.~\eqref{deriv} and \eqref{deriv2} are valid even in this case, and we have
\begin{equation} \begin{array}{ll}
\left. \frac{d}{d \lambda} {\mathcal E}_{K,Q} (\psi + \lambda \phi) \right|_{\lambda = 0}
= {\rm Re} \int_{\Omega} dr^3 ~  |\nabla \psi|^{q-2}  \psi \cdot {\bar \phi}
= {\rm Re} \int_{\Omega}  {\mathcal F}_{K,Q}(\psi) {\bar \phi} ~dr^3  =  {\rm Re} ({\mathcal F}_{K,Q}(\psi), \phi). 
\end{array} \end{equation}

\subsubsection{Nonlinear interaction depending on the parameter ${\tilde t}_5$} \label{ex3}
The nonlinear term represented using the Laplacian operator is considered.
We consider the energy represented by
\[{\mathcal E}_L(\psi) = \frac{1}{4}  \int_{\Omega} dr^3  |\psi|^2 \triangle |\psi|^2 ,  \]
where the corresponding Hamiltonian energy density is ${\mathcal H}_L (\psi) =  \frac{1}{4} |\psi|^2 \triangle |\psi|^2 $. 
By minimizing this energy (i.e., the Gateaux differential is considered), we obtain
\[ \begin{array}{ll}
{\mathcal F}_{L}(\psi) =  (\triangle |\psi|^{2}) \psi .
\end{array} \]
Indeed, for $\psi$ and $\phi$ in $W^{2,4}(\Omega)$,
\[  \begin{array}{ll}
 |\psi + \lambda \phi |^2 \triangle |\psi + \lambda \phi|^2 -  |\psi|^2 \triangle |\psi|^2 
 = \int_0^1 \frac{\partial}{\partial t} \{ |\psi + t \lambda \phi |^2 \triangle |\psi + t \lambda \phi|^2 \}~  dt \vspace{1.5mm}\\
\quad = 2 \lambda  \int_0^1 \big[ ( \triangle |\psi + t \lambda \phi|^2) ~ {\rm Re}( (\psi + t \lambda \phi) \cdot {\bar \phi} )
 + {\rm Re}  \triangle ((\psi + t \lambda \phi) \cdot {\bar \phi} )  |\psi + t \lambda \phi|^2  \big]  dt  \vspace{1.5mm}\\
\end{array} \]
is true, where
\[  \begin{array}{ll}
 ( \triangle |\psi + t \lambda \phi|^2) ~ {\rm Re}( (\psi + t \lambda \phi) \cdot {\bar \phi} ) ~\to~ 
 ( \triangle |\psi|^2) ~ {\rm Re}( \psi \cdot {\bar \phi} )  \quad  {\rm a.e.},
 \vspace{1.5mm}\\
  \triangle ( |\psi + t \lambda \phi| ~ \phi )  |\psi + t \lambda \phi|^2  ~\to~
  \triangle ( |\psi| \phi )~  |\psi|^2   \quad  {\rm a.e.} 
\end{array}  \]
are valid due to $\lambda \to 0$ ($\lambda > 0$).
We have
\[ \begin{array}{ll}
\left. \frac{d}{d \lambda} {\mathcal E}_L (\psi + \lambda \phi) \right|_{\lambda = 0}
= \frac{1}{2} \int_{\Omega} dr^3 ~ \{ ( \triangle |\psi|^2) ~  {\rm Re} ( \psi \cdot {\bar \phi} ) + {\rm Re} \triangle (\psi \cdot {\bar \phi}) ~ |\psi|^2  \}.  
\end{array} \]
If $\psi$ and $\phi$ are sufficiently smooth, by taking into account the boundary condition: the boundary conditions shown in Sec.~\ref{ex0}, we have
\[  \begin{array}{ll}
 \int_{\Omega}   \triangle ( |\psi| ~ \phi )  |\psi|^2   ~dr^3  =
- \int_{\Omega}   \nabla ( |\psi| ~ \phi ) \nabla |\psi|^2  ~dr^3  =
 \int_{\Omega}   \triangle |\psi|^2 ~ {\rm Re} (\psi ~{\bar \phi})  ~dr^3.
\end{array} \]
Consequently,
\begin{equation} \begin{array}{ll}
\left. \frac{d}{d \lambda} {\mathcal E}_L (\psi + \lambda \phi) \right|_{\lambda = 0}
=  {\rm Re} \int_{\Omega} dr^3 ~  \{  ( \triangle |\psi|^2) ~  \psi \cdot  {\bar \phi}   \}
= {\rm Re} \int_{\Omega}  {\mathcal F}_{L}(\psi) {\bar \phi} ~dr^3  =  {\rm Re} ({\mathcal F}_{L}(\psi), \phi).
\end{array} \end{equation}
Note that this interaction including differential operators cannot be bounded on $L^2(\Omega)$. \vspace{16mm} \\

\subsubsection{Nonlinear interaction arising from current (depending on the parameter ${\tilde t}_4$)} \label{ex4}
The momentum density ${\bm j}$, which is defined by ${\bm j} = - \frac{i}{2} ( {\bar \psi} \nabla \psi - \psi \nabla {\bar \psi})$, plays a significant role more than one physical quantity to be utilized to describe the interaction. 
For a master equation: $i \partial_t \psi = (1/2) (-\Delta + V) \psi$ in a Hilbert space $L^2(\Omega)$,  
\[ \begin{array}{ll}
\frac{\partial}{\partial t} ({\bar \psi} \psi) =  (\partial_t {\bar \psi}) \psi + {\bar \psi} (\partial_t \psi) \vspace{1.5mm}\\ 
\quad  = \frac{1}{2} (i(-\Delta + V) {\bar \psi}) \psi + \frac{1}{2} {\bar \psi} (-i(-\Delta + V) \psi \vspace{1.5mm}\\ 
\quad  =  \frac{i}{2}((-\Delta {\bar \psi}) \psi  - {\bar \psi} (-\Delta \psi)) + \frac{i}{2}((V {\bar \psi}) \psi - {\bar \psi} (V \psi)).  \vspace{1.5mm}\\ 
\end{array} \]
If $V$ is a self-adjoint operator in $L^2(\Omega)$, then $ \frac{\partial}{\partial t} ({\bar \psi} \psi) 
 =  (i/2)((-\Delta {\bar \psi}) \psi  - {\bar \psi} (-\Delta \psi)) = (i/2) \nabla \cdot ({\bar \psi} \nabla \psi - \psi \nabla {\bar \psi})$ follows, and 
\begin{equation} \frac{\partial}{\partial t} ({\bar \psi} \psi) +  \nabla \cdot {\bm j} = 0 \end{equation}
is obtained.
This equation, which means the conservation of total particle density (represented by $\int dr^3 |\psi|^2$), is known as the continuity equation.
That is, the momentum density ${\bm j}$ plays a role of current.

We consider the energy represented by
\[{\mathcal E}_J(\psi) =  \int_{\Omega} dr^3 {\bm j}^2 = \frac{(-i)^2}{4} \int_{\Omega} dr^3  ( {\bar \psi} \nabla \psi - \psi \nabla {\bar \psi})^2,  \]
where the corresponding Hamiltonian energy density is ${\mathcal H}_J (\psi) = - \frac{1}{4} ({\bar \psi} \nabla \psi - \psi \nabla {\bar \psi})^2 $. 
By minimizing this energy (i.e., the Gateaux differential is considered), we obtain
 \[{\mathcal F}_J(\psi) =  - 2 i \{ 2 (\nabla \cdot {\bm j}) \psi  + {\bm j} \cdot  \nabla \psi \}.\]
Indeed, for $\psi$ and $\phi$ in $W^{1,4}(\Omega)$,
\[ \begin{array}{ll}
 \{ (\overline{ \psi + \lambda \phi}) \nabla (\psi + \lambda \phi)  - (\psi + \lambda \phi) \nabla  (\overline {\psi + \lambda \phi}) \}^2 -  ( {\bar \psi} \nabla \psi - \psi \nabla {\bar \psi})^2  \vspace{1.5mm}\\
\quad = \int_0^1 \frac{\partial}{\partial t} \{ (\overline {\psi + t \lambda \phi}) \nabla (\psi + t \lambda \phi)  - (\psi + t \lambda \phi) \nabla  (\overline{ \psi + t \lambda \phi}) \}^2   dt  \vspace{1.5mm}\\
\quad =  \int_0^1 dt ~ \big[  2\{ (\overline {\psi + t \lambda \phi}) \nabla (\psi + t \lambda \phi)  - (\psi + t \lambda \phi) \nabla  (\overline{ \psi + t \lambda \phi}) \}    \vspace{1.5mm}\\
\qquad  \frac{\partial}{\partial t}  \{ (\overline{ \psi + t \lambda \phi}) \nabla (\psi + t \lambda \phi)  - (\psi + t \lambda \phi) \nabla  (\overline{ \psi + t \lambda \phi}) \} \big]   \vspace{1.5mm}\\
\quad = 2 \lambda   \int_0^1 dt ~ \big[  \{ (\overline{ \psi + t \lambda \phi}) \nabla (\psi + t \lambda \phi)  - (\psi + t \lambda \phi) \nabla  (\overline{ \psi + t \lambda \phi}) \}    \vspace{1.5mm}\\
\qquad \{ {\bar \phi} \nabla (\psi + t \lambda \phi) + (\overline{ \psi + t \lambda \phi}) \nabla \phi  - \phi \nabla  (\overline{ \psi + t \lambda \phi})   - (\psi + t \lambda \phi) \nabla {\bar \phi} \} \big]. 
\end{array} \]
It follows that
\[ \begin{array}{ll}
 \frac{{\mathcal E}_J(\psi + \lambda \phi) - {\mathcal E}_J(\psi)}{\lambda}
= - \frac{1}{2} \int_{\Omega} dr^3 ~ \int_0^1 dt~  \big[  \{ (\overline{ \psi + t \lambda \phi}) \nabla (\psi + t \lambda \phi)  - (\psi + t \lambda \phi) \nabla  (\overline{ \psi + t \lambda \phi}) \}    \vspace{1.5mm}\\
\qquad \{ {\bar \phi} \nabla (\psi + t \lambda \phi) + (\overline{ \psi + t \lambda \phi}) \nabla \phi  - \phi \nabla  (\overline{ \psi + t \lambda \phi})   - (\psi + t \lambda \phi) \nabla {\bar \phi} \} \big]. 
\end{array} \]
Due to $\lambda \to 0$ ($\lambda > 0$),
\[  \begin{array}{ll} 
  - \{ (\overline{\psi + t \lambda \phi}) \nabla (\psi + t \lambda \phi)  - (\psi + t \lambda \phi) \nabla  (\overline{ \psi + t \lambda \phi}) \} 
 ~\to~ 
  - \{{\bar \psi} \nabla \psi  - \psi \nabla {\bar \psi} \} = 2i {\bm j}
 \quad {\rm a.e.}, \vspace{2.5mm}\\
 \{{\bar \phi} \nabla (\psi + t \lambda \phi) + (\overline{ \psi + t \lambda \phi}) \nabla \phi  - \phi \nabla  (\overline{ \psi + t \lambda \phi})   - (\psi + t \lambda \phi) \nabla {\bar \phi} \} 
 ~\to~
 {\bar \phi} \nabla \psi + {\bar \psi} \nabla \phi  - \phi \nabla {\bar \psi}  - \psi \nabla {\bar \phi}  
 \quad {\rm a.e.}
 \end{array} \] 
are valid.
If $\psi$ and $\phi$ are sufficiently smooth, we have
\begin{equation} \label{inbyp} \begin{array}{ll}
\left. \frac{d}{d \lambda} {\mathcal E}_J (\psi + \lambda \phi) \right|_{\lambda = 0}
= - \int_{\Omega} dr^3 ~ (i {\bm j}) \{ (\nabla \psi) {\bar \phi} + {\bar \psi}  \nabla \phi   -  (\nabla  {\bar \psi}) \phi  - \psi \nabla{\bar \phi}   \} \vspace{1.5mm}\\
 =  -  \int_{\partial \Omega} i ( {\bm j} \cdot {\bm \nu}) {\bar \psi} \phi ~ dS + \int_{\partial \Omega}  i ( {\bm j} \cdot {\bm \nu}) \psi {\bar \phi} ~ dS
 \vspace{1.5mm}\\
\quad + \int_{\Omega} dr^3 ~ \{ -i{\bm j}  (\nabla \psi) {\bar \phi} +i (\nabla \cdot {\bm j}){\bar  \psi} \phi  +i {\bm j}  (\nabla {\bar \psi}) \phi  +i{\bm j} (\nabla {\bar \psi})   \phi  -i (\nabla \cdot {\bm j}) \psi {\bar \phi}   -i{\bm j} (\nabla \psi) {\bar \phi}   \}  \vspace{1.5mm}\\
 = -  \int_{\partial \Omega} i ({\bm j} \cdot {\bm \nu}) {\bar \psi} \phi ~ dS + \int_{\partial \Omega}  i ({\bm j} \cdot {\bm \nu}) \psi {\bar \phi} ~ dS  \vspace{1.5mm}\\
\quad +  \int_{\Omega} dr^3 ~ \{ - 2i{\bm j}  (\nabla \psi) {\bar \phi}  + 2 i {\bm j}  (\nabla {\bar \psi}) \phi  +i (\nabla \cdot{\bm j})  {\bar \psi} \phi   -i (\nabla \cdot {\bm j}) \psi {\bar \phi}   \} \vspace{1.5mm}\\
 = - \int_{\partial \Omega}  i ({\bm j} \cdot {\bm \nu}) {\bar \psi} \phi ~ dS + \int_{\partial \Omega}  i ({\bm j} \cdot {\bm \nu}) \psi {\bar \phi} ~ dS 
 +  \int_{\Omega} dr^3 ~ \{  - 4 {\rm Re} ( i{\bm j}  (\nabla \psi) {\bar \phi} )   - 2 {\rm Re} ( i (\nabla \cdot {\bm j})  \psi {\bar \phi})   \},
\end{array} \end{equation}
where $\nu$ means the outward normal vector for the $\partial \Omega$.
By taking into account the boundary condition: the boundary conditions shown in Sec.~\ref{ex0}, the integral on the boundary surface cancels.
Therefore
\[ \begin{array}{ll}
\left. \frac{d}{d \lambda} {\mathcal E}_J (\psi + \lambda \phi) \right|_{\lambda = 0}
=  {\rm Re}  \int_{\Omega} dr^3 [ -2i \{ 2 {\bm j}  (\nabla \psi) {\bar \phi}    +  (\nabla \cdot {\bm j})  \psi {\bar \phi}  \}]   
=  ({\mathcal F}_J(\psi), \phi) ,
\end{array} \]
where it is very important to note that the second term of the right hand side of the following equation:
\begin{equation} \label{missing}  ({\mathcal F}_J (\psi), \phi) =   {\rm Re}  \int_{\Omega}  \frac{4}{i} \{ {\bm j}  (\nabla \psi) {\bar \phi}    +  (\nabla \cdot {\bm j})  \psi {\bar \phi}  \}  ~ dr^3  -  {\rm Re}  \int_{\Omega}  \frac{2}{i} 
 (\nabla \cdot {\bm j})  \psi {\bar \phi} ~ dr^3
 \end{equation}
is missing in the standard formalism shown in the Appendix B of Ref.~\cite{engel-brink}.
This term arises from the careful treatment of the integration by parts (see~Eq.\eqref{inbyp}).
Note that this interaction including differential operators cannot be bounded on $L^2(\Omega)$.

\subsubsection{Effective Hamiltonian}
The corresponding effective Hamiltonian for the Hamiltonian density shown in Eq.~\eqref{hamdensexam} is obtained as
\begin{equation} \begin{array}{ll}
{\mathcal H}_{\rm eff}({\bm r}) =  - {\tilde t}_1 2 \triangle +  {\tilde t}_2 4 \rho 
 + {\tilde t_3} 2(1 + {\tilde \alpha})  \rho^ {\tilde \alpha} \vspace{1.5mm}\\
\qquad + {\tilde t}_4 \{ - 2 \nabla \cdot ( |\psi|^2  \nabla \psi)  + 2  |\nabla \psi|^{2} \psi + 2i \{ 2 {\bm j}  (\nabla \psi) +  (\nabla \cdot {\bm j})  \psi \} 
  \}  +{\tilde t}_5 4 \triangle  \rho  \vspace{2.5mm}\\
\quad = 2 [  - {\tilde t}_1  \triangle + 2 {\tilde t}_2  \rho 
 + (2 +  {\tilde \alpha}) {\tilde t_3} \rho^{1 + {\tilde \alpha}} \vspace{1.5mm}\\
\qquad + {\tilde t}_4 \{ -  \nabla \cdot ( |\psi|^2  \nabla \psi)  +   |\nabla \psi|^{2} \psi + i \{ 2 {\bm j}  (\nabla \psi) +  (\nabla \cdot {\bm j})  \psi \} 
  \}  + 2 {\tilde t}_5  \triangle  \rho ],    
\end{array} \end{equation}
where we should pay attention to the coefficients with signs.
Two points should be noticed: as is seen in ${\tilde t}_1$, if we obtain the coefficient in the effective Hamiltonian as $- \hbar^2/2m$, the corresponding coefficient $t_1$ in the Hamiltonian density should be $\hbar^2/4m$ (it requires the modification of the standard choice of coefficients contained in the Hamiltonian density~\cite{vautherin,engel-brink}); as is already pointed out, it is necessary to subtract $ 2i  (\nabla \cdot {\bm j})  \psi$ (compared to the standard parametrization shown in the Appendix B of Ref.~\cite{engel-brink}) in order to have a complete set of terms arising from $ {\bm j}^2$ in the Hamiltonian density.
For the former point, it is useful to remember that
\[
\int_{\Omega}  \tau dr^3  ~ = ~  \int_{\Omega} - (\triangle \psi) {\bar \psi} dr^3; \qquad   \tau =| \nabla \psi |^{2}
\]
does not follow from the variational principle, but from the integration by parts with a suitable boundary condition.
Although the representation of the standard Hamiltonian density is not correct~\cite{vautherin,engel-brink}, the standard effective Hamiltonian based on such a Hamiltonian density is exactly the same as the effective Hamiltonian obtained here (except for the terms arising from $ {\bm j}^2$).
For the latter point, the Galilean invariance has to be broken if the missing term is not included (for the Galilean invariance, see also the corresponding discussion in Sec.~\ref{sub2}).

In this section the discussion has been developed in $L^p$ spaces, where $p$ is not necessarily equal to 2.
Indeed, $p ~(\ne 2)$ is necessary to consider nonlinear problems; e.g., for $\psi \in L^2(\Omega)$, $|\psi|^2$ is included in $L^1(\Omega)$ at least, and therefore we are not sure if $|\psi|^2 \psi$ is included even in $L^1(\Omega)$.  
Note again that $L^p (\Omega)$ are not Hilbert spaces, if $p$ is not equal to 2.
In such situations ($p \ne 2$), the inner product is not equipped, so that we cannot discuss the orthogonality of functions and so on.
Here is a difficulty of considering nonlinear problems.  \vspace{16mm} \\

\section{Remarks on the additional forces} \label{sec3}
\subsection{The Coulomb force}
Only protons hold charge, so that they interact by the Coulomb force.
Charge is assumed to be equal to the probability distribution of proton (cf. form factor). 
The Coulomb force consists of the direct and the exchange parts.
The direct part of the Coulomb energy is represented by 
\[ E_{C}^{dir} = \frac{e^2}{2} \int \int dr_i^3  d r_j^3  \frac{ \rho_p({\bm r}_i)  \rho_p({\bm r}_j) }{|{\bm r}_i-{\bm r}_j|},  \]
using the proton density, where the corresponding Hamiltonian density is equal to $\frac{e^2}{2}  \int  dr_j^3  \frac{ \rho_p({\bm r}_i)  \rho_p({\bm r}_j) }{|{\bm r}_i-{\bm r}_j|}$.
On the other hand, the exchange part of the Coulomb energy is approximated by means of the Slater approximation \cite{slater} as
\[ E_{C}^{ex} = - \frac{3e^2}{4} \left(\frac{3}{\pi} \right)^{1/3}  \int dr_i^3  \rho_p({\bm r}_i)^{4/3},  \]
where the corresponding Hamiltonian density is equal to $ - (3e^2/4) ({3/\pi})^{1/3}  \rho_p({\bm r}_i)^{4/3}$.
Note that the exchange part of the Coulomb force has the similar form as the term with the coefficient $t_3$, so that the treatment of obtaining the corresponding part of the effective Hamiltonian is similar to the cases with $E_{{\tilde t}_3}$.

\begin{figure} \label{fig1} 
\includegraphics[width=6cm]{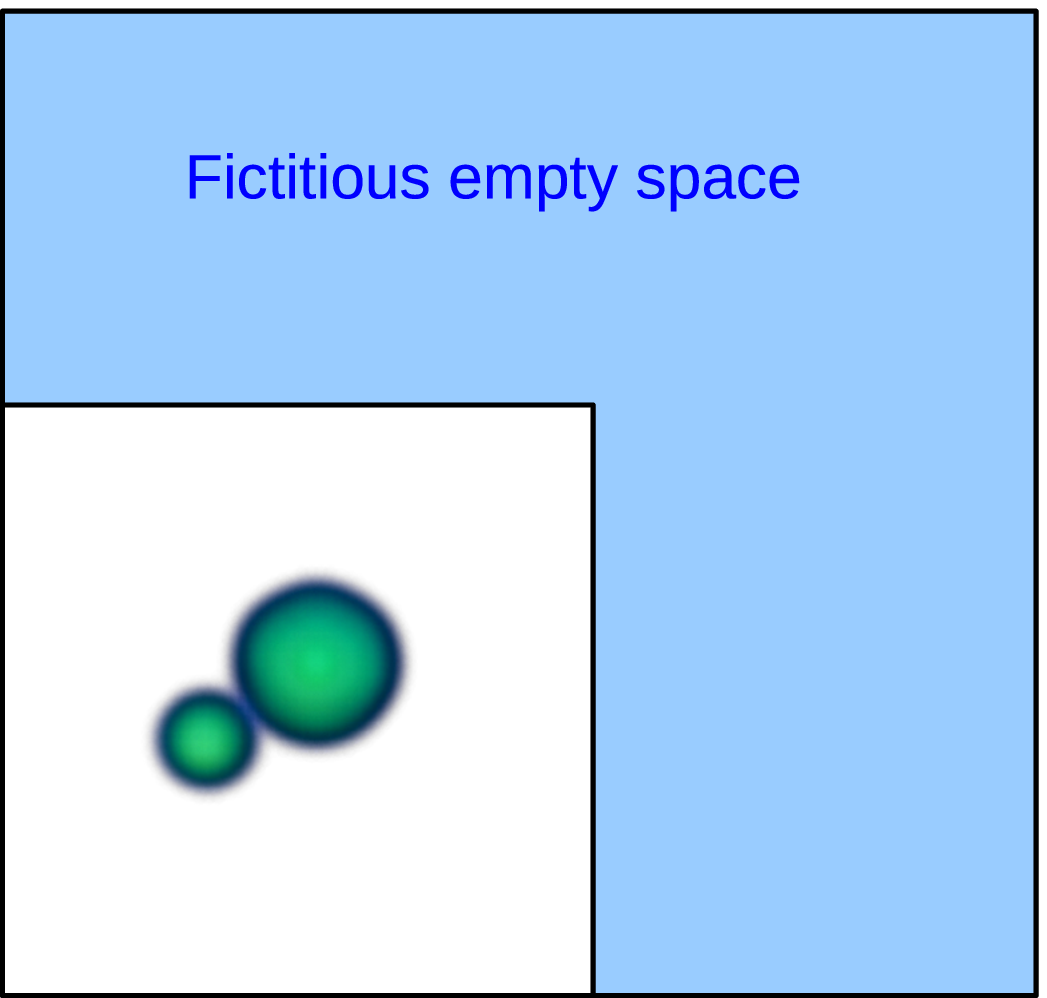} \\
\caption{
An actual numerical calculation of the Coulomb interaction is presented in the coordinate space when the periodic boundary condition is imposed.
A white-coloured square means the computational cell.
The solution is obtained by two fast Fourier transform operations in the enlarged region with periodic boundary condition.
This treatment realizes the isolated charge distribution in the computational cell.}
\end{figure}

Let us move on to the numerical calculation of the Coulomb force. 
Because of the long-range property of the Coulomb interaction, the periodic boundary condition is not necessarily appropriate, but the potential has to go to zero at infinity (``isolated charge distribution").
In practice this case is solved usually in one of two ways, either by obtaining boundary values from a multi-pole expansion or by Fourier techniques embedding the computational cell in one large of empty cells (Fig.~2).
In the periodic case being useful for astrophysical situation, the jellium approximation is used corresponding to a constant background density of electrons cancelling the total charge. \vspace{16mm} \\

\subsection{Pairing force}
The pairing interaction is an important ingredient of quantum many-body systems.
This interaction combines two fermions into one boson, so that condensation can take place as a new feature.
For example the superconductivity follows from the pairing interaction.

In order to introduce the pairing interaction to the density functional theory, it is necessary to have a functional representation of the pairing-interaction field.
There have been proposed two methods of introducing the pairing interaction in many-nucleon systems~\cite{Ring-Schuck}; one is the HFB (Hartree-Fock-Bogoliubov) approach, and the other is the BCS (Bardeen, Cooper, and Schrieffer) approach. 

Following the review article~\cite{stone-reinhard}, here we introduce two pairing-field of BCS type:
\begin{equation} \label{pair1}  \begin{array}{ll} E_{\rm pair} = \sum_q \frac{v_q}{4} \int dr^3~ \chi_q({\bm r})^2 \end{array}  \end{equation}
and 
\begin{equation} \label{pair2}  \begin{array}{ll} E_{\rm pair}^{DD} = \sum_q \frac{v_{0,q}}{4}~\int dr^3~ \left[ \left\{ 1- \left( \frac{\rho({\bm r})}{\rho_c} \right)^{\gamma} \right\} \chi_q({\bm r})^2 \right], \end{array} \end{equation}
where
\[ \chi_q ({\bm r}) = \sum_{ {\hat \alpha} \in q} w_ {\hat \alpha} u_{\hat \alpha} v_{\hat \alpha} |\psi_{\hat \alpha} ({\bm r}) |^2\]
denotes the pairing density with the phase-space weight $w_{\hat \alpha}$, occupation amplitude $v_{\hat \alpha}$, and non-occupation amplitude $u_{\hat \alpha} = \sqrt{1-v_{\hat \alpha}^2} $.
In Eqs.~\eqref{pair1} and \eqref{pair2}, $v_q$ and $v_{0,q}$ are strength parameters, and $\rho_0$ is the nuclear saturation density, typically $\rho_0 = 0.16$~fm$^3$.
In contrast to Eq.~\eqref{pair1}, Eq.~\eqref{pair2} additionally includes the density dependence.
The surface profile of the pairing interaction is controlled by the corresponding parameter $\gamma$, whose standard value is equal to 1. 
Note that the strong pairing takes place near the nuclear surface, so that Eq.~\eqref{pair2} is expected to describe the pairing field better than Eq.~\eqref{pair1}.
The corresponding Hamiltonian densities for Eq.~\eqref{pair1} and Eq.~\eqref{pair2} are equal to $ \sum_q \frac{v_q}{4} \chi_q({\bm r})^2$ and $ \sum_q \frac{v_{0,q}}{4}~ [ \{ 1- ( \frac{\rho({\bm r})}{\rho_c} )^{\gamma} \} \chi_q({\bm r})^2 ]$, respectively.
As is readily seen, the treatment of obtaining the corresponding parts of the effective Hamiltonian are similar to the cases with $E_{{\tilde t}_2}$ and $E_{{\tilde t}_3}$.
 \vspace{16mm} \\

\section{Summary} \label{sec4}
A whole process of deriving the effective interaction in many-nucleon systems has been shown.
What is presented in this chapter is a kind of modelling: the modelling of interacting many-nucleon systems.
In particular we have illustrated the appearance of both nonlinearity and differential operators in this formalism.
In the application of variational principle, based on the functional analytic methods, we have presented a treatment of the Gateaux differential in some generalized situations.

Phenomenological density-dependent force has been treated.
The exchange part of the Coulomb force and the pairing force have fractional powers of the density, so that the density-dependent property is also true to these two forces.
Such a fractional power dependence reasonably appears in the physics treating finite quantum systems.
Indeed, as far as finite quantum systems are concerned, the existence of $\rho$ term in energy implies the emergence of $\rho^{2/3}$ term in energy due to the surface effects.

Two unknown features have been found.
First, as is shown in the first term of the right hand side of Eq.~\eqref{hamdens}, the coefficient of the kinetic energy part of the Hamiltonian density is equal to $\hbar^2/4m$.
This fact requires the modification of the standard choice~\cite{vautherin,engel-brink} of coefficients contained in the Hamiltonian density.
Second, as is shown in Eq.~\eqref{missing}, we have pointed out that there is a missing term in the standard derivation of the effective interaction arising from ${\bm j}^2$ in the Hamiltonian density.
This fact requires the modification of the standard choice ~\cite{engel-brink} of coefficients contained in the effective interaction.
The correct treatment of this term is necessary not only to hold the Galilean invariance, but also to clarify the time-odd contribution to the stationary and non-stationary states of many-nucleon system.
\vspace{16mm} \\

This work was supported by the Helmholtz Alliance HA216/EMMI.
One of the authors (Y. I.) expresses his gratitude to Prof. Emeritus. Dr. Hiroki Tanabe (Department of Mathematics, Osaka University), who made many valuable comments with respect to the mathematical rigorous treatment of density functional.  \vspace{16mm} \\

\section{Appendix -differentiability -} \label{appdx}
Let $z$ be a notation for complex variable:
\[ \begin{array}{ll}
 \frac{\partial}{\partial z} = \frac{1}{2} \left(  \frac{\partial}{\partial x} - i \frac{\partial}{\partial y}  \right), \quad
 \frac{\partial}{\partial {\bar z}} = \frac{1}{2} \left(  \frac{\partial}{\partial x} + i \frac{\partial}{\partial y}  \right). 
\end{array} \]
We consider a function $f(z)$ satisfying $f(z) = f(x,y)$ and $z=x+iy$.
Here let us assume that $f(x,y)$ is a differentiable function of the two real variables $(x,y)$.
According to the Taylor's theorem, 
\[ \begin{array}{ll}
f(z +\omega) -  f(z) = f(x +\xi,y+\eta) -  f(x,y)
= \frac{\partial f(x,y)}{\partial x} \xi +  \frac{\partial f(x,y)}{\partial y} \eta + \cdots,  
\end{array} \]
where $\omega = \xi +\eta$ and
\[ \begin{array}{ll}
 \frac{\partial}{\partial x} =  \frac{\partial}{\partial z} + \frac{\partial}{\partial {\bar z}}, \quad
 \frac{\partial}{\partial y} =  i \left(  \frac{\partial}{\partial z} - \frac{\partial}{\partial {\bar z}}  \right). 
\end{array} \]
Therefore
\begin{equation} \label{refeq}  \begin{array}{ll}
f(z +\omega) -  f(z)
= \left( \frac{\partial f(z)}{\partial z} + \frac{\partial f(z)}{\partial {\bar z}} \right) \xi 
+ i \left( \frac{\partial f(z)}{\partial z} - \frac{\partial f(z)}{\partial {\bar z}} \right) \eta + \cdots \vspace{1.5mm} \\
\quad =  \frac{\partial f(z)}{\partial z} (\xi + i \eta)  + \frac{\partial f(z)}{\partial {\bar z}}  (\xi - i \eta)
 + \cdots  \vspace{1.5mm} \\
\quad =  \frac{\partial f(z)}{\partial z} \omega  + \frac{\partial f(z)}{\partial {\bar z}} {\bar \omega} + \cdots.    
\end{array} \end{equation}
In particular $f$ is Fr\'{e}chet differentiable if $\partial f / \partial {\bar z} = 0$ (i.e., $f$ is holomorphic).
However, even though $f$ is not holomorphic,
\[ \begin{array}{ll}
\lim_{\lambda \to 0} \frac {f(z + \lambda \omega) -  f(z)}{\lambda} =
 \frac{\partial f(z)}{\partial z} \omega  +  \frac{\partial f(z)}{\partial {\bar z}} {\bar \omega}
 \end{array}  \]
is valid to real $\lambda$, if $f$ is differentiable.
In this situation it is possible to consider Gateaux differential:  
\[ Df(z)(\omega)  =
 \frac{\partial f(z)}{\partial z} \omega  +  \frac{\partial f(z)}{\partial {\bar z}} {\bar \omega}.  \]
In particular if $f$ is real-valued, the Gateaux differential is reduced to
\begin{equation} \label{gatea}  Df(z)(\omega)  =
2 {\rm Re} \left(  \frac{\partial f(z)}{\partial z} \omega \right),  \end{equation}
where note that $\partial f(z)/\partial z$ and $\partial f(z)/\partial {\bar z}$ are complex-conjugate.
In the case of Sec.~\eqref{ex0}, the right hand side of Eq.~\eqref{gatea} corresponds to
\[ \frac{\lambda}{d \lambda} {\mathcal E}(\psi + \lambda \phi)|_{\lambda = 0}  = - {\rm Re} \int_{\Omega} dr^3 | \nabla \psi|^{p-2} \nabla \psi \nabla {\bar \phi},  \]
but the limit ($\lambda \to 0$) cannot exist if $\lambda$ is not real (on the other hand, $\lambda$ should be complex with respect to finding the optimal condition in the complex Banach spaces).
Indeed, according to Eq. \eqref{refeq}, the limit of
\[ \begin{array}{ll}
\frac { {\mathcal E}(\psi + \lambda \phi) -  {\mathcal E}(\psi)}{\lambda} =
 \frac{\partial {\mathcal E}(\psi)}{\partial \psi} \phi  +  \frac{\partial {\mathcal E}(\psi)}{\partial {\bar \psi}} \frac{\bar \lambda}{\lambda} {\bar  \phi} +  \cdots
 \end{array}  \]
due to $\lambda \to 0$ ($\lambda$ is not real) exists only when ${\mathcal E}$ is holomorphic.
Note that ${\mathcal E}$ shown in Sec.~\eqref{ex0} is not holomorphic. 

In summary a mapping from a complex Banach space to a complex Banach space is Fr\'{e}chet differentiable, only if the function with complex variables is holomorphic.
On the other hand, the real-valued functions are not holomorphic except for the constant.
It provides a reason why we have lengthy treatments of the Gateaux differential as demonstrated in Sec.~\ref{sub3}.
\end{document}